\RequirePackage{fix-cm}
\documentclass[smallextended]{svjour3}       %
\smartqed  %
\usepackage{graphicx}
\journalname{Theoretical and Computational Fluid Dynamics}

\usepackage{amsmath} %
\usepackage{graphicx} %

\usepackage{subcaption}
\usepackage{lipsum}
\usepackage{xcolor}

\def\fade{0}
\def\period{\mathcal{P}}
\def\zMInf{a_{\infty^-}}
\def\zPInf{a_{\infty^+}}

\usepackage{tikz,pgfplots,pstool}

\usetikzlibrary{calc, 
				decorations.markings, 
				decorations.text,
				decorations.pathreplacing,
				shapes.multipart,
				positioning,
				arrows.meta,
				arrows,
				matrix,
				patterns,
				shadows,
				hobby}

\tikzset{
    side by side/.style 2 args={
        line width=2pt,
        #1,
        postaction={
            clip,postaction={draw,#2}
        }
    },
    circle node/.style={
        circle,
        draw,
        fill=white,
        minimum size=1.3cm
    }
}

\if\fade1
\usetikzlibrary{fadings}
\tikzfading[name=fade out,
inner color=transparent!0,
outer color=transparent!100]
\fi

\tikzset{test/.style={
    postaction={
        decorate,
        decoration={
            markings,
            mark=at position \pgfdecoratedpathlength-0.5pt with {\arrow[blue,line width=#1] {}; },
            mark=between positions 0 and \pgfdecoratedpathlength-0pt step 0.5pt with {
                \pgfmathsetmacro\myval{multiply(divide(
                    \pgfkeysvalueof{/pgf/decoration/mark info/distance from start}, \pgfdecoratedpathlength),100)};
                \pgfsetfillcolor{white!\myval!black};
                \pgfpathcircle{\pgfpointorigin}{#1};
                \pgfusepath{fill};}
}}}}

\def\skCircDetColThick#1#2#3#4{
	\begin{scope}[shift={#1}]
		\draw[line width = \lw, #4,fill = confgrey] (0,0) circle (#2);
		\draw[fill=black] (0,0) circle (.01);
		\draw[latex-latex] (0,-.05)--(#2,-.05) node[midway, above] {$q_#3$};
		\node [left, below] {$\delta_#3$};
		\node[right = 3pt,above = 3pt] at ($#2*({cos(30)},{sin(30)})$) {$C_{#3}$};
	\end{scope}
}

\def\skCircBlankBThick#1#2#3{
	\begin{scope}[shift={#1}]
		\draw[line width = \lw, #3,fill = confgrey] (0,0) circle (#2);
	\end{scope}
}

\def\roundPolyO#1#2{
	\begin{scope}[shift={#1}]
		\draw[line width = \lw, blue,scale=#2,fill=confgrey,rotate = 15] (-1,-1)--(-1,1)--(1,1)--(1,-1)--cycle;
	\end{scope}
}

\def\roundPolyI#1#2{
	\begin{scope}[shift={#1}]
		\draw[line width = \lw, red,scale=#2,fill=confgrey,rotate = 0,rounded corners = 2mm] (-1,-.8)--(-.4,.8)--(.8,-.8)--cycle;
	\end{scope}
}

\def\roundPolyII#1#2{
	\begin{scope}[shift={#1}]
		\draw[line width = \lw, green,scale=#2,fill=confgrey, rotate = 0,rounded corners = 2mm] (-.8,-1)--(-1,.8)--(-0,.4)--(.8,1.1)--(.9,-.8)--cycle;
	\end{scope}
}

\tikzset{
    mark position/.style args={#1(#2)}{
        postaction={
            decorate,
            decoration={
                markings,
                mark=at position #1 with \coordinate (#2);
            }
        }
    }
}

\def\myCanCircTripBranchCaseIThick{
	\path [path fading = circle with fuzzy edge 15 percent, fill=confgrey, even odd rule] (0,0) circle (\zeroRad) (0,0) circle (1.3*\zeroRad);
	\draw[line width = \lw, blue,fill= confcol] (0,0) circle (\zeroRad);
	
	\node[right = 8pt, above = 8pt] at ($\zeroRad*({cos(30)},{sin(30)})$) {};
	
	\node at (-1,1.75) {$D_\zeta$};
	\skCircBlankBThick{(\delIx,\delIy)}{\radI}{red}
	\skCircBlankBThick{(\delIIx,\delIIy)}{\radII}{green}
	
	\draw[*-*,line width = \lw,dashed,branchCol] (-1.8,1) .. controls (-1,-1) and (0,.5) .. (1.75,-.8) node[black,pos=0,above] {$\zPInf$} node [black, pos=1,below] {$\zMInf$};
	
}

\def\myCanCircTripBranchCaseIIThick{
	\path [path fading = circle with fuzzy edge 15 percent, fill=confgrey, even odd rule] (0,0) circle (\zeroRad) (0,0) circle (1.3*\zeroRad);
	\draw[line width = \lw, blue,fill= confcol] (0,0) circle (\zeroRad);
	
	\node[right = 8pt, above = 8pt] at ($\zeroRad*({cos(30)},{sin(30)})$) {};
	
	\node at (-1,1.75) {$D_\zeta$};
	\skCircBlankBThick{(\delIx,\delIy)}{\radI}{red}
	\skCircBlankBThick{(\delIIx,\delIIy)}{\radII}{green}
	
	\draw[*-,line width = \lw,dashed,branchCol] (-1,.5) .. controls (.5,-1) and (1,0) .. ($({\zeroRad*cos(30)},-{\zeroRad*sin(30)})$) node[black,pos=0,above] {$a_{\infty}$};
	
}

\def\myCanCircTripBranchCaseIIIThick{
	\path [path fading = circle with fuzzy edge 15 percent, fill=confgrey, even odd rule] (0,0) circle (\zeroRad) (0,0) circle (1.3*\zeroRad);
	\draw[ line width = \lw, blue,fill= confcol] (0,0) circle (\zeroRad);
	
	\node[right = 8pt, above = 8pt] at ($\zeroRad*({cos(30)},{sin(30)})$) {};
	
	\node at (-1,1.75) {$D_\zeta$};
	
	\draw[*-,line width = \lw,dashed,branchCol] (\delIx,\delIy) .. controls (-.5,.5) and (-1.5,-1.5) .. ($(-{\zeroRad*cos(30)},-{\zeroRad*sin(30)})$);
	
	\skCircBlankBThick{(\delIx,\delIy)}{\radI}{red}
	\skCircBlankBThick{(\delIIx,\delIIy)}{\radII}{green}
	
}

\def\myCanCircQuadDetailedThick{
	\path [path fading = circle with fuzzy edge 15 percent, fill=confgrey, even odd rule] (0,0) circle (\zeroRad) (0,0) circle (1.3*\zeroRad);
	\draw[line width = \lw, blue,fill= confcol] (0,0) circle (\zeroRad);
	
	\node[right = 8pt, above = 8pt] at ($\zeroRad*({cos(30)},{sin(30)})$) {$C_0$};
	
	\node at (-1.5,1.5) {$D_\zeta$};
	\skCircDetColThick{(\delIx,\delIy)}{\radI}{1}{red}
	\skCircDetColThick{(\delIIx,\delIIy)}{\radII}{2}{green}
	\skCircDetColThick{(\delIIIx,\delIIIy)}{\radIII}{3}{magenta}
	
}

\def\myShiftUp#1{\raisebox{1ex}}
\def\myShiftDown#1{\raisebox{-2.5ex}}

\def\myPathTextAbove#1#2#3#4{
\draw [-LaTeX,thick,postaction={decorate,
               decoration={
                         raise=1ex,
                         text along path,
                         text align={center},
                         text={%
                                     |\color{black}| {#1} }
                              }
                        }
	  ] #2 to [bend left=#4] #3;
}

\def\myPathTextLeftAbove#1#2#3#4{
	\draw [LaTeX-,thick,postaction={decorate,
		decoration={
			raise=1ex,
			text along path,
			text align={center},
			text={%
				|\color{black}| {#1} }
		}
	}
	] #2 to [bend left=#4] #3;
}

\usetikzlibrary{decorations.text,fit,chains,calc,shapes.geometric,intersections}

\newlength{\myww}
\newlength{\myhh}

\definecolor{themecolour}{rgb}{0.64, 0.76, 0.68}
\definecolor{mygray}{rgb}{0.7, 0.7, 0.7}
\definecolor{tablegray}{rgb}{0.75, 0.75, 0.75}
\definecolor{confcol}{RGB}{255, 255, 255}
\definecolor{camblue}{RGB}{108, 172, 228}
\definecolor{camred}{RGB}{213, 0, 50}
\definecolor{camnavy}{RGB}{0, 60, 113}
\definecolor{camgreen}{RGB}{114, 180, 49}

\definecolor{myblue}{rgb}{0 ,  0.4470 , 0.7410}
\definecolor{myorange}{rgb}{0.8500,    0.3250,    0.0980}
\definecolor{myyellow}{rgb}{0.9290,    0.6940,    0.1250}
\definecolor{mypurple}{rgb}{ 0.4940,    0.1840,    0.5560}
\definecolor{myred}{rgb}{     0.6350 ,   0.0780 ,   0.1840}
\definecolor{mygreen}{rgb}{         0.4660  ,  0.6740   , 0.1880}

\definecolor{branchCol}{rgb}{0, .8, .8}

\definecolor{prsared}{RGB}{ 219,25,73}
\definecolor{prsablue}{RGB}{ 4,146,210}

\definecolor{confblue}{rgb}{0.75, 0.72, 0.95}
\definecolor{confgreen}{rgb}{0.76, 1, 0.74}
\definecolor{confgrey}{rgb}{0.9, .9, 0.9}

\definecolor{cvblue}{rgb}{0.22,0.45,0.70}%

\definecolor{matlab1}{rgb}{0,0.4470,0.7410}
\definecolor{matlab2}{rgb}{0.8500,0.3250,0.0980}
\definecolor{matlab3}{rgb}{0.9290,0.6940,0.0980}
\definecolor{matlab4}{rgb}{0.4940,0.1840,0.5560}
\definecolor{matlab5}{rgb}{0.4660,0.6740,0.1880}
\definecolor{matlab6}{rgb}{0.3010,0.7450,0.9330}
\definecolor{matlab7}{rgb}{0.6350,0.0780,0.1840}
\def\imageFolder{images}

\def\zeroRad{3}
\def\delIx{1}
\def\delIy{1}
\def\radI{1}
\def\delIIx{0}
\def\delIIy{-1.5}
\def\radII{.75}
\usepackage{amsfonts}
\def\radIII{.5}
\def\delIIIx{-1.5}
\def\delIIIy{0}

\def\lw{1}
\usepackage{esint}
\newcommand{\citep}{\cite}
\makeatletter
\graphicspath{{/images/}}
\makeatother

\usetikzlibrary{external}
\usepackage{hyperref}
\hypersetup{
	colorlinks,
	linkcolor={red!50!black},
	citecolor={blue!50!black},
	urlcolor={blue!80!black}
}

\begin{document}

\title{A calculus for flows in periodic domains%
}

\author{Peter J. Baddoo         \and
        Lorna J. Ayton %
}

\institute{%
              Department of Applied Mathematics and Theoretical Physics \\
              University of Cambridge\\
Wilberforce Road\\
Cambridge\\
CB3 0WA\\
              \email{baddoo@damtp.cam.ac.uk}     
}

\date{Received: 1st December 2019 }

\maketitle

\begin{abstract}
We present a constructive procedure for the calculation of 2-D potential flows in periodic domains with multiple boundaries per period window. The solution requires two steps: (i) a conformal mapping from a canonical circular domain to the physical target domain, and (ii) the construction of the complex potential inside the circular domain. All singly periodic domains may be classified into three distinct types: unbounded in two directions, unbounded in one direction, and bounded. In each case, we relate the target periodic domain to a canonical circular domain via conformal mapping and present the functional form of prototypical conformal maps for each type of target domain. We then present solutions for a range of potential flow phenomena including flow singularities, moving boundaries, uniform flows, straining flows and circulatory flows. By phrasing the solutions in terms of the transcendental Schottky--Klein prime function, the ensuing solutions are valid for an arbitrary number of obstacles per period window. Moreover, our solutions are exact and do not require any asymptotic approximations.
\end{abstract}

\def\e{\textrm{e}}
\def\d{\textrm{d}}
\def\i{\textrm{i}}

\section{Introduction} \label{Sec:intro}
Spatially periodic domains arise in almost every area of fluid dynamics. For example, in geophysical fluid dynamics, the surface of the earth can be projected onto a periodic grid through a conformal projection \citep{Lee1976}, and porous media can be represented as arrays of periodically spaced pores \citep{Chapman1992}. Modern aerofoil designs also exploit periodic features in the form of drag-reducing riblets \citep{Ehrenstein1996} and noise-reducing serrations \citep{Ayton2018}. In other aerospace applications, individual turbomachinery stages may be modelled as a periodic ``cascade'' of aerofoils \citep{Schmid2017}, thereby permitting both aerodynamic \citep{Robinson1956,Baddoo2018c} and aeroacoustic \citep{Evers2002,Peake1997} analyses. In other applications, superhydrophobic surfaces are often manufactured with patterned longitudinal periodic arrays of ridges \citep{Kirk2018}. In summary, the accurate and versatile mathematical modelling of flows through periodic domains is essential for elucidating the underlying physical mechanisms associated with such flows. In this article we provide a constructive procedure for the calculation of such flows, i.e. a calculus for flows in periodic domains.

Typically, the solution of 2-D potential flow problems require two steps \citep{Fokas2003}: (i) a conformal mapping from a (multiply connected) canonical circular domain to the physical periodic target domain of interest, and (ii) the solution of the potential flow problem inside the circular domain. Due to the invariance of Laplace's equation under conformal mappings, these two steps combine to solve the full flow problem in the physical domain \citep{Katz2009}. We adopt an analogous approach in our work and accordingly there are two quantities to obtain in our calculus: the conformal mapping and the potential flow solution. Historically, conformal mappings have been restricted to simply connected, non-periodic domains. The archetypal example is the Joukowski mapping \citep{Joukowski1910} which relates the unit disc to a Joukowski-type aerofoil. The extension of potential theory to multiply connected domains came about in the early 2000s through the identification of the \emph{Schottky--Klein prime function} \citep{Baker1897} as a fundamental object associated with multiply connected domains (\cite{Crowdy2005b,Crowdy2006,Crowdy2005,Crowdy2008,Crowdy2008a}). The prime function has also found relevance in fluid mechanics problems since \cite{Crowdy2010} presented a ``new calculus of vortex dynamics'' to enable the calculation of 2-D potential flows in multiply connected domains. This present paper is a natural sequel of that work and we present an extension of \cite{Crowdy2010} to periodic domains. Similarly to \cite{Crowdy2010}, an appealing feature of our solutions is that they are valid for multiply connected domains i.e. multiple bodies per period window. In fact, using the prime function allows us to express the solutions in a consistent manner regardless of connectivity.

The authors were motivated towards this study by a need to calculate the potential flow through a cascade of aerofoils to analyse turbomachinery noise \citep{Baddoo2020Aero}. In previous work, they were able to find an asymptotic solution in the thin aerofoil limit, where the angle of attack and aerofoil aspect ratio are assumed to be small \citep{Baddoo2018c}. Whilst that solution offers physical insight in the form of asymptotic expansions, it was limited to a small class of aerofoil geometries, and only a single boundary per period window. Conversely, the solutions we present in this paper are valid for any geometry (provided the appropriate conformal mapping is available), and can account for multiple boundaries per period window.

The remainder of the article is arranged as follows. In section \ref{Sec:formulation} we present the mathematical formulation of our problem and introduce the key mathematical objects. In particular, we define the canonical circular domain in section \ref{Sec:circ} and the periodic target domain in section \ref{Sec:target}. In particular, we differentiate between three possible types of target domain. We then present the functional form of the conformal mappings from a circular domain to these three types in \mbox{section \ref{Sec:conf}}. In doing so, we introduce the Schottky--Klein prime function (\mbox{section \ref{Sec:SKP}}), which serves as an essential tool for constructing potential flows through multiply connected domains. We then proceed in \mbox{section \ref{Sec:sols}} by calculating the potential flow within the multiply connected circular domain. We devote attention to the cases of uniform flow, straining flow, and circulatory flows, which require special treatment due to the periodicity of the target domain. Finally, we summarise our results in \mbox{section \ref{Sec:conclusions}} and suggest applications and future lines of research.
\section{Mathematical formulation} \label{Sec:formulation} 
In this paper we seek to construct 2-D potential flow solutions in periodic domains. We now introduce the main mathematical objects used to develop our solutions.

\subsection{The canonical circular domain} \label{Sec:circ}
We now define the canonical circular domain, denoted by $D_\zeta$. We take $D_\zeta$ to be the interior of the unit disc with $M$ excised discs so that there are a total of $M+1$ boundary circles. The unit disc is labelled as $C_{0}$ and the excised discs as labelled as $\left\{C_{j} | j = 1, \dots , M \right\}$. The excised discs have centers $\left\{\delta_j | j = 1, \dots , M \right\}$ and radii $\left\{q_j | j = 1, \dots , M \right\}$. 
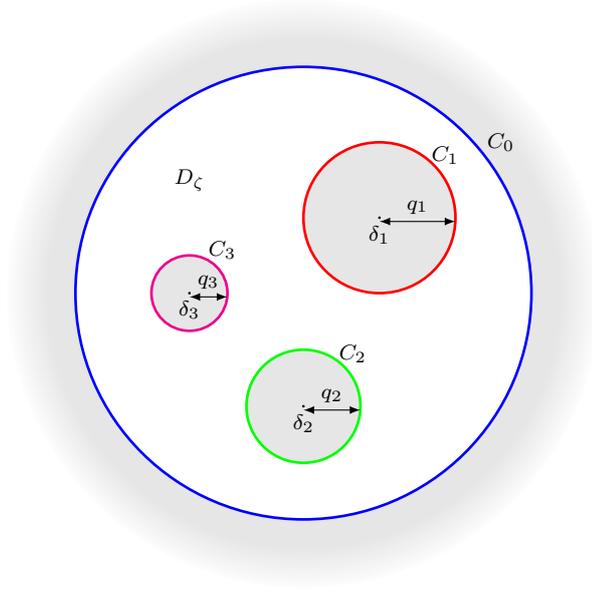
\begin{figure}
\centering
\begin{tikzpicture}[remember picture]

\myCanCircQuadDetailedThick

\end{tikzpicture}

\caption{A multiply connected circular domain $D_\zeta$ for $M=3$. $C_{0}$ denotes the unit circle and $\left\{C_{j}| j = 1,\cdots, 3\right\}$ denotes the boundaries of the
excised discs with centers $\delta_j$ and radii $q_j$. The grey colour denotes regions that are outside the domain of definition.}

\label{Fig:MCD}
\end{figure}
For example, in the simply connected case ($M=0$), the canonical circular domain is the unit disc. In the doubly connected case ($M=1$), the canonical circular domain may be taken to be a concentric annulus without loss of generality. A typical quadruply connected ($M=3$) circular domain is illustrated in figure \ref{Fig:MCD}. In general, every point on the circle $C_j$ satisfies
\begin{align}
|\zeta -\delta_j|^2 = (\zeta - \delta_j)(\overline{\zeta} - \overline{\delta}_j) = q_j^2. \label{Eq:mob1}
\end{align}
Moreover, every circle $C_j$ has an associated M\"obius map defined by
\begin{align}
\theta_j(\zeta) = \delta_j + \frac{q_j^2 \zeta}{1 - \overline{\delta}_j \zeta}.
\label{Eq:mobMap}
\end{align}
These M\"obius maps have a specific geometrical meaning. For $j>0$, we define $C_j^\prime$ to be the circle obtained by reflecting the circle $C_j$ in the unit disc $C_0$. Using \eqref{Eq:mob1}, it is possible to show that the image of $C_j^\prime$ in the M\"obius map $\theta_j$ is simply the circle $C_j$. Therefore, each M\"obius map $\theta_j$ represents the reflection of the disc $C_j$ in the unit circle $|\zeta| = 1$. 

The structure of the branches in the circular domain is dependent on the periodic target domain: in case I there are two interior branch points, in case II there is only one interior branch point, and in case III there are zero interior branch points although two interior circles are connected by a branch cut. This structure is illustrated on the left hand side of figure \ref{Fig:conformal}.

\subsection{The target domain} \label{Sec:target}
\begin{figure}
	\centering
	\begin{subfigure}[!h]{\linewidth}
		\centering
		\begin{tikzpicture}[remember picture, scale = .9]
		\begin{scope}[scale =.8, shift = {(-.55\linewidth,0)}]
		
		\myCanCircTripBranchCaseIThick
		
		\end{scope}

		\def\stag{3}
		\begin{scope}[xscale=.5, yscale = .5,shift = {(.4\linewidth,0)}, rotate = 90]

		\fill[path fading = north, white] (-6,-3) rectangle (-5,7);

		\node at (1.5,2.5) {$D_z$};
		
		\foreach \x in {-1,0,1,2}
		{
			\roundPolyO{(-3,\stag*\x)}{.8}
			\roundPolyI{(3,\stag*\x)}{1}
			\roundPolyII{(0,\stag*\x)}{1}
		}
		\foreach \x in {-1,0,1}
		{
			\draw[dashed,line width = \lw,shift={(0,\x*\stag+\stag/2)},domain=-6:6,samples = 100,smooth,variable=\xVar,branchCol] plot ({\xVar},{.1*sin(\xVar*120)});
		}
		
		\fill[white,path fading = west] (-6,-\stag-.01) rectangle (6,-\stag/2);
		\fill[white] (-6,-\stag/2-\stag) rectangle (6,-\stag);
		
		\fill[white,path fading = east] (-6,2*\stag+0.01) rectangle (6,\stag + \stag/2);
		\fill[white] (-6,\stag/2+2*\stag) rectangle (6,2*\stag);
		
		\fill[white,path fading = north] (-6-0.01,-\stag) rectangle (-5,2*\stag);
		\fill[white,path fading = south] ( 5,-\stag) rectangle ( 6+0.01,2*\stag);
		
		\draw[latex-] (-6.5,0)--(-5.5,0) node[pos=1,above] {\small$f(a_{\infty^-})$};
		\draw[-latex] (5,0)--(6,0) node[pos=0,below] {\small $f(a_{\infty^+})$};
		
		\end{scope}
		
		\myPathTextAbove{$z = f(\zeta)$}{(-2.2,1.75)}{(-1.2,1.75)}{20}
		
		\end{tikzpicture}
		\caption{Type I.} \label{Fig:typeI}
	\end{subfigure}
	
	\begin{subfigure}[!h]{\linewidth}
		\centering
		\begin{tikzpicture}[remember picture,scale =.9]
		\begin{scope}[scale = .8, shift = {(-.55\linewidth,0)}]
		
		\myCanCircTripBranchCaseIIThick
		
		\end{scope}

		\def\stag{3}
		\begin{scope}[xscale=.5, yscale = .5,shift = {(.4\linewidth,0)}, rotate = 90]
		
		\fill[path fading = north, white] (-6,-3) rectangle (-5,7);

		\foreach \x in {-1,0,1}
		{
			\draw[dashed,line width = \lw,shift={(0,\x*\stag+\stag/2)},domain=-4:6,samples = 100,smooth,variable=\xVar,branchCol] plot ({\xVar},{.1*sin(\xVar*120)});
		}

		\node at (1.5,2.5) {$D_z$};
		
		\foreach \x in {-1,0,1,2}
		{
			\roundPolyI{(3,\stag*\x)}{1}
			\roundPolyII{(0,\stag*\x)}{1}
			\draw[line width = \lw, blue,fill=confgrey, rounded corners = 10mm] (-4,-.5*\stag+\stag*\x)--(-2,\stag*\x)--((-4,.5*\stag+ \stag*\x);
			\fill[confgrey] (-6,-.5*\stag + \stag*\x) rectangle (-4, .5*\stag + \stag*\x);
		}
		
		\fill[white,path fading = west] (-6,-\stag-.01) rectangle (6,-\stag/2);
		\fill[white] (-6,-\stag/2-\stag) rectangle (6,-\stag);

		\fill[white,path fading = east] (-6,2*\stag+0.01) rectangle (6,\stag + \stag/2);
		\fill[white] (-6,\stag/2+2*\stag) rectangle (6,2*\stag);
		
		\fill[white,path fading = north] (-6-0.01,-\stag) rectangle (-5,2*\stag);
		\fill[white,path fading = south] ( 5,-\stag) rectangle ( 6+0.01,2*\stag);
		
		\draw[-latex] (5,0)--(6,0) node[pos=0,below] {\small $f(a_\infty) $};
		
		\end{scope}
		
		\myPathTextAbove{$z = f(\zeta)$}{(-2.2,1.75)}{(-1.2,1.75)}{20}
		\end{tikzpicture}
		\caption{Type II.} \label{Fig:typeII}
	\end{subfigure}
	
	\begin{subfigure}[!h]{\linewidth}
		\centering
		\begin{tikzpicture}[remember picture,scale = .9]
		\begin{scope}[scale = .8, shift = {(-.55\linewidth,0)}]
		
		\myCanCircTripBranchCaseIIIThick
		
		\end{scope}

		\def\stag{3}
		\begin{scope}[xscale=.5, yscale = .5,shift = {(.5\linewidth,0)}, rotate = 90]
		
		\fill[path fading = north, white] (-6,-3) rectangle (-5,7);
		
		\foreach \x in {-1,0,1}
		{
			\draw[dashed,line width = \lw,shift={(0,\x*\stag+\stag/2)},domain=-4:4,samples = 100,smooth,variable=\xVar,branchCol] plot ({\xVar},{.1*sin(\xVar*120)});
		}
		
		\node at (1.5,2.5) {$D_z$};
		
		\foreach \x in {0,1}
		{
			\roundPolyII{(0,\stag*\x)}{1}
			\draw[line width = \lw, blue,fill=confgrey,rounded corners = 10mm ] (-4,-.5*\stag+\stag*\x)--(-2,\stag*\x)--((-4,.5*\stag+ \stag*\x);
			\fill[confgrey] (-6,-.5*\stag + \stag*\x) rectangle (-4, .5*\stag + \stag*\x);
			\fill[confgrey] (6,-.5*\stag + \stag*\x) rectangle (4, .5*\stag + \stag*\x);
			\draw[line width = \lw, red,fill=confgrey] (4,-.5*\stag+\stag*\x)--(4,-.25*\stag+ \stag*\x)--(2.5,-.25*\stag+\stag*\x)--(2.5,.25*\stag+\stag*\x)--(4,.25*\stag+\stag*\x)--(4,.5*\stag+ \stag*\x);
		}
		
		\begin{scope}[shift = {(0,-\stag)}]
		
		\roundPolyII{(0,0)}{1}
		\draw[line width = \lw, blue,fill=confgrey,rounded corners = 10mm] (-4,-.5*\stag)--(-2,0)--((-4,.5*\stag);
		\fill[confgrey] (-6,-.5*\stag) rectangle (-4, .5*\stag);
		\fill[confgrey] (6,-.5*\stag) rectangle (4, .5*\stag);
		\draw[line width = \lw, red,fill=confgrey] (4,-.5*\stag)--(4,-.25*\stag)--(2.5,-.25*\stag)--(2.5,.25*\stag)--(4,.25*\stag)--(4,.5*\stag);
		\end{scope}
		
		\begin{scope}[shift = {(0,2*\stag)}]
		
		\roundPolyII{(0,0)}{1}
		\draw[line width = \lw, blue,fill=confgrey,rounded corners = 10mm] (-4,-.5*\stag)--(-2,0)--(-4,.5*\stag);
		\fill[confgrey] (-6,-.5*\stag) rectangle (-4, .5*\stag);
		\fill[confgrey] (6,-.5*\stag) rectangle (4, .5*\stag);
		\draw[line width = \lw, red,fill=confgrey] (4,-.5*\stag)--(4,-.25*\stag)--(2.5,-.25*\stag)--(2.5,.25*\stag)--(4,.25*\stag)--(4,.5*\stag);
		
		\end{scope}

		\fill[white,path fading = west] (-6,-\stag-.01) rectangle (6,-\stag/2);
		\fill[white] (-6,-\stag/2-\stag) rectangle (6,-\stag);

		\fill[white,path fading = east] (-6,2*\stag+0.01) rectangle (6,\stag + \stag/2);
		\fill[white] (-6,\stag/2+2*\stag) rectangle (6,2*\stag);
		
		\fill[white,path fading = north] (-6-0.01,-\stag) rectangle (-5,2*\stag);
		\fill[white,path fading = south] ( 5,-\stag) rectangle ( 6+0.01,2*\stag);
		
		\end{scope}
		
		\myPathTextAbove{$z = f(\zeta)$}{(-2.2,1.75)}{(-1.2,1.75)}{20}
		
		\end{tikzpicture}
		\caption{Type III.} \label{Fig:typeIII}
	\end{subfigure}
	
  \caption{The preimage and target domains in type I, II and III domains in the case $M=2$. The preimages of $\pm \i \infty$, if they exist in the target, are at $a_{\infty^\pm}$ or $a_{\infty}$. The branch cut is denoted by the \textcolor{branchCol}{light blue} curve.}
	\label{Fig:conformal}
\end{figure}
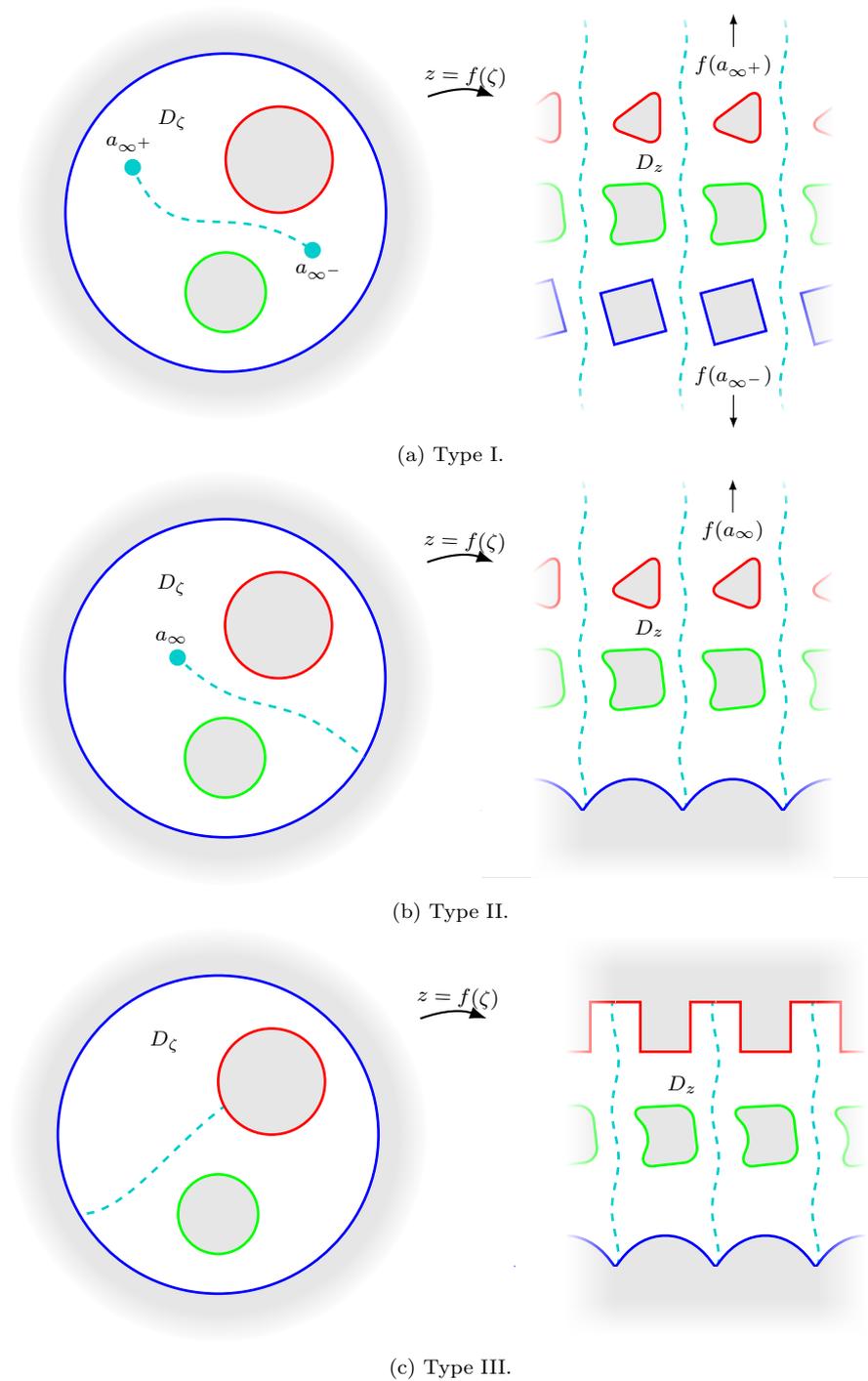

Every singly periodic domain belongs to one of three types,as illustrated on the right hand side of figure \ref{Fig:conformal}: type I, where the period window is unbounded in two directions; type II, where the period window is unbounded in only one direction; and type III, where the period window is bounded in every direction. We term the periodic domain of interest the ``\emph{target domain}". Moreover, each of these classes may be related to a canonical circular domain with an appropriate branch structure, as illustrated on the left hand side of figure \ref{Fig:conformal}. The circular domain is discussed in more detail in section \ref{Sec:circ}. The target domain consists of an arrangement of identical period windows with period $\period$. In this paper we assume that $\period$ is real, although complex periods may be obtained through an elementary rotation. The target domain consists of $M+1$ boundaries which we label $\left\{L_j\; | \; j = 0,\cdots, M \right\}$. We label the target domain $D_z$ and endow it with complex coordinate $z$.

\subsection{Periodic conformal mappings} \label{Sec:conf}
In the present work we consider mappings from multiply connected circular domains (labelled $D_\zeta$) to target period windows of single periodic domains (labelled $D_z$), as illustrated in figure \ref{Fig:conformal}. In this section we present the prototypical form of each type of mapping, along with simple examples to aid intuition. In section \ref{Sec:SC} we recapitulate a constructive formula for computing mappings to polygonal domains.

\subsubsection{Type I periodic conformal mappings}

Type I periodic geometries are unbounded in two directions; a typical type I geometry and its corresponding canonical circular domain are illustrated in figure \ref{Fig:typeI}. Accordingly, the mapping function $f(\zeta)$ must contain precisely two branch points in the circular domain $D_\zeta$: one that is mapped to $+\i\infty$, which we denote $a_{\infty^+}$, and another that is mapped to $-\i \infty$, which we label $a_{\infty^-}$. Moreover, these branch points must have equal and opposite coefficients, otherwise the mapping does not have the correct multi-valued structure. For period $\period$, we may express every type I mapping in the general form
\begin{align}
f(\zeta) = \frac{\period}{2 \pi \i} \log \left(\frac{\zeta - a_{\infty^+}}{\zeta - a_{\infty^-}}\right) + \tilde{f}(\zeta), \label{Eq:typeI}
\end{align}
where $\tilde{f}$ is analytic in $D_{\zeta}$. It is straightforward to check that winding clockwise around $\zeta = a_{\infty^\pm}$ yields an increase in $f$ of $\pm \period$ per loop. Type I mappings have found relevance in the study of superhydrophobic surfaces \citep{Crowdy2011}, where the mappings were used to derive frictional slip lengths for grooved surfaces. Additionally, these mappings have been applied to find analytic solutions for free boundary problems, including von K\'arm\'an streets of hollow vortices \citep{Crowdy2011a} for both simply connected and doubly connected domains, and arrays of bubbles in Hele--Shaw cells  \citep{Vasconcelos1993,Vasconcelos2015}

Perhaps the most simple type I periodic conformal mappings is
\begin{align}
f(\zeta)  = \frac{\period}{\pi} \left(\arctan\left(\frac{\zeta}{a_{\infty}}\right) - \arctan(\zeta a_{\infty})\right), \label{Eq:tanMap}
\end{align}
where we have taken $a_\infty = a_{\infty^+} = -a_{\infty^-}$ to be any point inside the unit disc. The mapping \eqref{Eq:tanMap}
 transplants the unit disc to a periodic array of horizontal slits with unit period. This mapping and its variants have been previously applied in turbomachinery studies to calculate the background potential flow through a cascade of flat plate aerofoils \citep{Robinson1956,Evers2002,Peake1995}. The approach presented herein allows more general studies to be conducted, using alternative flow phenomenon and higher connectivities.

Some new periodic slit maps for type I geometries are presented in appendix \ref{Ap:conf1}.
\subsubsection{Type II periodic conformal mappings} \label{Sec:conf2}
Type II periodic geometries are unbounded in one direction; a typical type II geometry and its corresponding canonical circular domain are illustrated in figure \ref{Fig:typeII}. Consequently, the mapping function $f(\zeta)$ must contain precisely one branch point in the circular domain $D_\zeta$. We may express every type II mapping in the general form
\begin{align}
f(\zeta) = \frac{\period}{2 \pi \i} \log \left(\zeta - a_{\infty}\right) + \tilde{f}(\zeta), \label{Eq:typeII}
\end{align}
where $\tilde{f}$ is analytic in $D_{\zeta}$. It is straightforward to check that winding clockwise around $\zeta = a_{\infty}$ yields an increase in $f$ of $\period$ per loop.

Type II mappings have previously been used to study the interaction of a vortex street with a shear flow in \cite{Crowdy2010a}, and free surface Euler flows in \cite{Crowdy2000}. In the simply connected case, \cite{Floryan1993} found an analogous form of the Schwarz--Christoffel formula to type II geometries, although the preimage domain in that case was the upper half-plane $\Im[\zeta]>0$.

The most simple type II periodic conformal mapping is 
\begin{align}
f(\zeta) & = \frac{\period}{2 \pi \i} \log \left(\frac{\zeta - a_{\infty}}{|a_\infty| (\zeta - 1/\overline{a_\infty})}\right), \label{Eq:logMap}
\end{align}
for any $a_\infty$ inside the unit disc. The mapping \eqref{Eq:logMap} transplants the unit disc to the real line in periodically repeated slits.

A new periodic slit map for type II geometries is presented in appendix \ref{Ap:conf2}.
\subsubsection{Type III periodic conformal mappings} \label{Sec:conf3}
Type III periodic geometries are bounded in both directions; a typical type III geometry and its corresponding canonical circular domain are illustrated in figure \ref{Fig:typeIII}. Consequently, the mapping function $f(\zeta)$ cannot contain any branch points in the circular domain $D_\zeta$. Instead, branch points are located at $\infty$ and at some $\gamma$ which we take to be inside the circle $C_1$. Therefore, the branch cut passes through the boundary circle $C_0$ and the boundary circle $C_1$. Accordingly, type III mappings may only exist when there is at least one excised circle ($M>0$). For period $\period$, we may express every type III mapping in the general form
\begin{align}
f(\zeta) = \frac{\period}{2 \pi \i} \log \left(\zeta - \gamma\right) + \tilde{f}(\zeta), \label{Eq:typeIII}
\end{align}
where $\gamma$ is located inside $C_1$ and $\tilde{f}$ is analytic in $D_{\zeta}$. It can be verified that winding clockwise around $C_0$ and $C_1$ results in an increase in $f$ of  $\period$ per loop. The main difference between the type III mapping \eqref{Eq:typeIII} and the type II mapping \eqref{Eq:typeII} is that the preimage of infinity is now located \emph{inside} one of the boundary circles.

Type III mappings have been applied to study steady capillary waves on an annulus \citep{Crowdy1999}, and to derive effective slip lengths for superhydrophobic surfaces \citep{Crowdy2017}. Additionally, Floryan \cite{Floryan1985a} derived the Schwarz--Christoffel formula for type III geometries, although the preimage domain was a horizontal channel $0<\Im[\zeta]<h$. The most straightforward type III mapping relates the concentric annulus to a periodic channel. Such a mapping takes the functional form
\begin{align*}
f(\zeta) &= \frac{\period}{2 \pi \i } \log(\zeta).
\end{align*}
When the annulus is of interior radius $q$, $C_0$ is mapped to the real axis and $C_1$ is mapped to the upper wall of the channel at a height of $\period\log(q)/(2 \pi \i)$. 

A new periodic slit map for type III geometries is presented in appendix \ref{Ap:conf3}.

\subsubsection{The Schottky--Klein prime function} \label{Sec:SKP}
The primary mathematical object in this paper is the ``Schottky--Klein prime function''. In particular, we will use this ``prime function'' to construct both the conformal mappings from  the circular domain to the target domain \emph{and} the complex potential in the circular domain. Whilst the prime function is well known in the context of Abelian functions \citep{Baker1897}, its relevance in the context of fluid dynamics problems has only been elucidated relatively recently \citep{Crowdy2010,Crowdy2005,Crowdy2008a}. The prime function is a transcendental analytic function associated with a particular canonical circular domain, such as that illustrated in \ref{Fig:MCD}. For brevity, we suppress the dependence of the prime function on the conformal moduli ($q_j$ and $\delta_j$) and write it as a bivariate function $\omega(\zeta, \alpha)$. 

For example, in the case $M=0$ where the canonical circular domain is the unit disc, the prime function is simply defined as
\begin{align*}
\omega(\zeta,\alpha)=\zeta - \alpha.
\end{align*}
In the doubly connected case ($M=1$), the canonical circular domain is the annulus 
\begin{align*}
\zeta = r \e^{\i \theta} ,\qquad \qquad q<r<1,\,\; 0<\theta<2 \pi.
\end{align*}
and the prime function may be written as
\begin{align*}
\omega(\zeta, \alpha)  = -\frac{\alpha}{C^2} P(\zeta/\alpha ,q), 
\end{align*}
where
\begin{align*}
P(\zeta,q) &= \left(1- \zeta\right) \prod_{k=1}^{\infty} \left(1 - q^{2k} \zeta \right) \left( 1- q^{2k} \zeta^{-1}\right),\\
C &= \prod_{k=1}^\infty (1 - q^{2k}).
\end{align*}
In the more general multiply connected case, the prime function may be expressed as the infinite product
\begin{align}
\omega(\zeta,\alpha) = (\zeta-\alpha) \prod_{\theta \in \Theta^{\prime \prime}} \frac{(\zeta - \theta(\alpha))(\alpha - \theta(\zeta))}{(\zeta - \theta(\zeta))(\alpha - \theta(\alpha))}, \label{Eq:skpDef}
\end{align}
where $\Theta^{\prime \prime}$ represents the Schottky group (which is the collection of all M\"obius maps $\theta_j$ defined in \eqref{Eq:mobMap}) excluding the identity and all inverses \citep{Crowdy2006}. It is possible to prove several salient properties of the prime function using the definition \eqref{Eq:skpDef}. Firstly, $\omega(\zeta,\alpha)$ is analytic everywhere inside $D_\zeta$ with a simple zero at $\zeta = \alpha$. Secondly, the prime function is skew-symmetric so
\begin{align*}
\omega(\zeta, \alpha) = - \omega(\alpha, \zeta),
\end{align*}
and its conjugate satisfies 
\begin{align*}
\overline{\omega} \left(\frac{1}{\zeta}, \frac{1}{\alpha}\right) & = -\frac{1}{\zeta \alpha} \omega(\zeta, \alpha),
\end{align*}
where the conjugate function is defined by $\overline{\omega}(\zeta,\alpha) = \overline{\omega(\overline{\zeta},\overline{\alpha})}$. Finally, the prime function transforms under the M\"obius maps $\theta_j$ as
\begin{align*}
\frac{\omega(\theta_j(\zeta),\alpha_1)}{\omega(\theta_j(\zeta),\alpha_2)} & = \beta_j(\alpha_1,\alpha_2) \frac{\omega(\zeta,\alpha_1)}{\omega(\zeta,\alpha_2)},
\end{align*}
where the functions $\beta_j$ are defined in \cite{Crowdy2006}.

In practice, the poor convergence properties of \eqref{Eq:skpDef} mean that it is rarely advisable to use this product definition in numerical computations. Moreover, it is not even guaranteed that the infinite products in \eqref{Eq:skpDef} converge for all choices of multiply connected domains. Alternatively,  recent algorithms presented by \cite{Crowdy2016a} provide a rapid and reliable scheme with which to compute the prime function, and a numerical implementation in \textsc{Matlab} is available at \url{https://github.com/ACCA-Imperial/SKPrime}. The approach of \cite{Crowdy2016a} is to express the prime function as the solution to a boundary value problem which can be solved with standard numerical methods. This approach supersedes the previous method of \cite{Crowdy2007d} where the prime function was expanded into a Fourier--Laurent expansion about the centers of the excised circles.

Further details on the prime function may be found in the review article \cite{Crowdy2016a}, and a monograph will be released soon \citep{CrowdyBook}.

\subsubsection{The periodic Schwarz--Christoffel formula} \label{Sec:SC}
Although the extended Riemann mapping theorem guarantees the existence of a conformal mapping between two domains of the same connectivity, it is often difficult to construct such a mapping in practice. Schwarz--Christoffel (S--C) formulae are useful tools for generating such mappings, as they represent constructive tools that furnish a conformal mapping to a desired target domain. Typically, S--C mappings provide a conformal mapping between a canonical domain (taken to be circles in the present work) and a polygonal domain, although there are extensions available for polycircular arc domains \citep{Crowdy2012a}, gear-like regions \citep{Goodman1960}, and curved regions \citep{Henrici1986}. Although the mappings are explicit, every S--C mapping is subject to a family of accessory parameters that must, in general, be determined numerically. A great deal of work has been devoted to solving this ``parameter problem'', and a comprehensive review is available in \cite{Driscoll2002}. In particular, the \textsc{Matlab} program \texttt{sc-toolbox} (\url{https://github.com/tobydriscoll/sc-toolbox}) allows the rapid computation of S--C mappings through the use of several novel numerical algorithms \citep{Driscoll1996,Driscoll2005}. 

Historically, Schwarz--Christoffel mappings were typically restricted to simply connected domains. A major advance came about in the early 2000s when two groups of researchers independently extended the S--C mapping formulae to consider multiply connected domains \citep{Delillo2004,Crowdy2005b,Crowdy2007}. The latter has an advantage over the former insofar as the mapping formula is written explicitly in terms of the aforementioned Schottky--Klein prime function. Further work has been done to solve the parameter problem in the multiply connected domains \citep{Kropf2012}. Recent work \cite{Baddoo2019c} has further extended the original S--C mapping to permit target domains that are periodic. Similarly to other work by Crowdy \citep{Crowdy2005b,Crowdy2007}, the mapping formula is phrased in terms of the Schottky--Klein prime function. Consequently, the formula is valid for any number of objects per period window. The mapping formula is given by (6.3) in \cite{Baddoo2019c} as
\begin{align}
f(\zeta) = A \int_1^\zeta S_{P} (\zeta^\prime) \prod_{j=0}^M \prod_{k=1}^{n_j} \left[\omega(\zeta^\prime,a_k^{(j)}) \right]^{\beta^{(j)}_{k}} \d \zeta^\prime+ B,
\label{intro:eq4}
\end{align}
where $\beta^{(j)}_k$ is the $k$-th turning angle on the $j$-th circle and $a_j^{(j)}$ are the pre-images of the vertices, which must generally be determined numerically. The constant $A$ represents a scaling and rotation, and $B$ represents a translation. The definition of the function $S_P$ is dependent on the type of mapping. In type I mappings, $S_P$ possesses simple poles at $a_{\infty^\pm}$ so that, when integrated, there are branch points at $a_{\infty^\pm}$. The original S--C mapping is recovered in the limit where $a_{\infty^\pm} \rightarrow a$ and the two branch points coalesce to form a simple pole at $\zeta=a$, thereby rending the map non-periodic. Conversely, in type II mappings $S_P$ only possesses a single pole (at $a_\infty$) whereas in type III mappings $S_P$ contains no poles in $D_\zeta$.

Now that we have presented the functional forms of periodic conformal mappings, we present our procedure for constructing the potential flow in the circular domain.

\section{Constructing the complex potential} \label{Sec:sols}
In this section we adapt the work of \cite{Crowdy2010} to enable the calculation of complex potentials for flows in periodic domains with multiple objects per period window. An incompressible and irrotational flow possesses a complex potential $w=\phi+\i \psi$ where $\phi$ and $\psi$ are the velocity potential and streamfunction respectively. Additionally, the complex potential is harmonic, so that
\begin{align*}
\nabla^2_z w = 0, \qquad \qquad z \in D_z.
\end{align*}
We write the complex potential in the $\zeta$-plane as $W(\zeta) = w(z(\zeta))$.

 Constructing the complex potential for a simply connected domain is relatively straightforward upon use of the Milne-Thomson circle theorem \citep{Batchelor1999}. However, problems in multiply connected domains are not subject to the same analysis, and the complex potential must be phrased in terms of the Schottky--Klein prime function. In this section, we present the complex potential for flow singularities (including point vortices, source-sink pairs, vortex spirals and doublets), flows induced by moving boundaries, uniform flows, higher-order flows, and circulatory flows. The analytic solutions for singularities and moving boundaries are essentially equivalent to the non-periodic analysis by \cite{Crowdy2010}, and do not require modification. Nevertheless, these solutions have not previously been applied to periodic domains and therefore we reiterate the original solutions by Crowdy. Conversely, the solutions for uniform flows, higher-order flows and circulatory flows must be modified to account for the periodicity of the domain. Each type of periodic geometry must be considered individually, although the solutions are all valid for arbitrary connectivities.

\subsection{Singularities}
Singularities embedded in a potential flow are frequently employed to model a host of physical phenomena. In particular, point vortices are commonly used to represent discretised vorticity, which finds relevance in geophysical fluid dynamics \citep{Southwick2015}, unsteady aerodynamics \citep{Darakananda2019}, and aeroacoustics \citep{Howe2003a}. Accordingly, the dynamics of point vortices has been described as a ``classical mathematics playground'' by \cite{Aref2007}. The advantage of modelling vorticity as point vortices is that the vorticity equation is replaced with a system of ordinary differential equations, which are usually far easier to solve numerically. The trajectories of more general classes of singularities -- sources and sinks, doublets etc. -- can also be computed using the approach advocated by \cite{LlewellynSmith2011}. In this section we recapitulate the complex potential for different types of singularities in multiply connected domains.
\subsubsection{Complex potential for point vortices, $W_V$} \label{Sec:vort}
A fundamental object in constructing potential flows in multiply connected domains is the hydrodynamic Green's function, $G_0$. The Green's function satisfies
	\begin{align}
	\nabla^2 G_0(\zeta,\alpha) & = \delta(\zeta-\alpha) , 
	\end{align}
for $\zeta \in D_{\zeta}$. Consequently, $G_0$ represents the complex potential induced by a unit strength point vortex at $\alpha$. For a circular domain of arbitrary connectivity, such as that illustrated in figure \ref{Fig:MCD}, \cite{Crowdy2005} showed that the hydrodynamic Green's function takes the form
\begin{align}
G_0(\zeta,\alpha) & = \frac{1}{2 \pi \i} \log \left(\frac{\omega(\zeta,\alpha)}{\left| \alpha\right| \omega(\zeta, 1/\overline{\alpha})}\right).\label{Eq:G0}
\end{align}
Furthermore, $G_0$ produces a circulation of $-1$ around $C_0$ and zero circulation around $C_j$,  $j = 1, \cdots, M$ . The circulations around each circle may be changed by introducing the \emph{modified} hydrodynamic Green's function
\begin{align}
G_j(\zeta,\alpha) & = \frac{1}{2 \pi \i} \log \left(\frac{\omega(\zeta,\alpha)}{\left| \alpha\right| \omega(\zeta, \theta_j(1/\overline{\alpha}))}\right). \label{Eq:modGreen}
\end{align}
The modified hydrodynamic Green's function $G_j$ produces $+1$ circulation around $C_j$ and zero circulation around $C_i$, for $i = 0, \cdots j-1, j+1, \cdots M$. We will use this fact later to construct flows with specified circulations around each object, thereby satisfying the Kutta condition.

Since the problem is linear, we may construct the complex potential induced by $n$ vortices of strength $\kappa_k$ located at $\alpha_k$ in the circular domain by writing
\begin{align}
W_{V}(\zeta; \mathbf{\alpha}) & = \sum_{k=1}^n \kappa_k G_{j_k}(\zeta, \alpha_k) \notag \\
& = \frac{1}{2 \pi \i }\sum_{k=1}^n \kappa_k \log \left(\frac{\omega(\zeta, \alpha_k)}{\left| \alpha_k\right| \omega(\zeta, \theta_{j_k} \left(1/ \overline{\alpha_k}\right))}\right). \label{Eq:vort1}
\end{align}
In many applications each object must have a specific circulation. Consequently, we may place point vortices at one of the preimages of infinity to alter the circulations around each body. For example, we may remove the circulation around every object by placing appropriately tuned vortices at, for example, $a_{\infty^-}$, resulting in
\begin{align}
W_{V}(\zeta;\mathbf{\alpha}, a_{\infty^-}) & = \sum_{k=1}^n \kappa_k \left(G_{j_k}(\zeta, \alpha_k) - G_{j_k}(\zeta, a_{\infty^-}) \right) \notag \\
& = \frac{1}{2 \pi \i }\sum_{k=1}^n \kappa_k \log \left(\left|\frac{a_{\infty^-}}{ \alpha_k} \right| \cdot \frac{\omega(\zeta, \alpha_k)}{\omega(\zeta, a_{\infty^\pm})} \cdot\frac{ \omega(\zeta, \theta_{j_k} \left(1/ \overline{a_{\infty^-}}\right))}{\omega(\zeta, \theta_{j_k} \left(1/ \overline{\alpha_k}\right))}\right). \label{Eq:vort2}
\end{align}

\subsubsection{Complex potential for a source-sink pair, $W_{P}$}
The complex potential for a source-sink pair of strength $m$ is given by \cite{Crowdy2013a} as
\begin{align}
W_P(\zeta;\alpha,\beta) = \frac{m}{2 \pi} \log \left(\frac{\overline{\alpha} \beta}{\left|\alpha \beta\right|} \cdot \frac{\omega(\zeta,\alpha) \omega(\zeta,\overline{\alpha}^{-1}}{\omega(\zeta, \beta) \omega(\zeta, \overline{\beta}^{-1})}\right), \label{Eq:ssPair}
\end{align}
where the source is located at $\alpha$ and the sink is located at $\beta$ in the circular domain. Additionally, $W_P$ induces zero circulation around every boundary.
\subsubsection{Complex potential for a vortex spiral, $W_{S}$}
The complex potential for vortex spirals at $\alpha$ and $\beta$ in a circular domain are obtained by combining  a source-sink pair \eqref{Eq:ssPair} with two point vortices:
\begin{align}
W_S(\zeta;\alpha,\beta) &= \frac{m}{2 \pi} \log \left(\frac{\overline{\alpha} \beta}{\left|\alpha \beta\right|} \cdot \frac{\omega(\zeta,\alpha) \omega(\zeta,\overline{\alpha}^{-1})}{\omega(\zeta, \beta) \omega(\zeta, \overline{\beta}^{-1})}\right) \\
&+ \frac{\Gamma}{2 \pi \i} \log\left(\frac{|\beta|}{|\alpha|} \cdot \frac{\omega(\zeta, \alpha)\omega(\zeta,\overline{\beta}^{-1}}{\omega(\zeta,\beta)\omega(\zeta, \overline{\alpha}^{-1})}\right). \label{Eq:spiral}
\end{align}
The complex potential $W_S$ produces zero circulation around every boundary.
\subsubsection{Complex potential for a doublet, $W_D$}
The complex potential for a doublet of unit strength at angle $\lambda$ at  $\zeta = \beta$ in a circular domain is given by \cite{Crowdy2010} as
\begin{align*}
W_D(\zeta;\lambda) &= 2 \pi \i \left[\e^{\i \lambda} \frac{\partial G_0}{\partial \overline{\alpha}}- \e^{-\i \lambda} \frac{\partial G_0}{\partial {\alpha}}\right]_{\alpha = \beta}
\end{align*}
It is possible to show that $W_D$ has a simple pole at $\zeta = \alpha$ such that 
\begin{align}
W_D(\zeta) \sim \frac{\e^{-\i \lambda}}{\zeta - \alpha}, \label{Eq:doubAsymp}
\end{align}
as $\zeta \rightarrow \alpha$.

\subsection{Complex potential for moving boundaries, $W_M$}
In many applications, the boundaries in periodic domains are moving. In this case, the typical no-flux boundary condition requires modification. In particular, the kinematic boundary condition states that the normal velocity of fluid on a rigid surface must move at the same velocity as that surface. Consequently, we write 
\begin{align}
\mathbf{u}\cdot \mathbf{n} &= \mathbf{U}_j \cdot \mathbf{n} \qquad \qquad \textnormal{on } L_j, \label{Eq:KinBCVec}
\end{align}
where $\mathbf{n}$ represents the normal direction, $\mathbf{u}$ represents the fluid velocity, and $\mathbf{U}_j$ represents the prescribed velocity of that boundary. In terms of the complex potential in the circular domain, \eqref{Eq:KinBCVec} corresponds to
\begin{align*}
	\Im \left[W_{M}(\zeta) \right] &= T_j(\zeta,\overline{\zeta}),  & \zeta \in  C_j, \;\; j = 0, \cdots, M
\end{align*}
where $T_j$ are functions relating to the specific motion of each body. For example, for rigid body motions, $T_j$ takes the form (see appendix \ref{Ap:rigid})
\begin{align*}
T_j(\zeta,\overline{\zeta}) & =\Im \left[ \dot{\bar{c}}_j(z(\zeta)-c_j) - \i \frac{\dot{\theta}_j}{2} \left| z(\zeta)-c_j\right|^2 \right]+d_j,
\end{align*}
where $c_j$ is a point in the object $L_j$ and $\theta_j$ is the angle of rotation of $L_j$ around $c_j$, and $d_j$ are constants that satisfy a compatibility condition to be defined below.

In other words, the problem is now to find an analytic function, $W_M$, such that the imaginary part of ${W_M}$ equals $T_j(\zeta,\bar{\zeta})$ on the boundary of each circle $C_j$. This is a form of the \textit{modified Schwarz problem}, the solution to which has been given by \cite{Crowdy2008} for a circular domain of arbitrary connectivity as
\begin{align}
W_M(\zeta) &= \frac{1}{2 \pi} \ointctrclockwise_{\partial D_\zeta} T_j\left(\zeta^\prime, \overline{\zeta^\prime}\right) \left( \d\log(\omega\left(\zeta^\prime,\zeta \right)) + \d \log\left( \overline{\omega} \left(\overline{\zeta^\prime}, \left. 1 \middle/ \zeta \right. \right)\right)\right) + E, \label{Eq:schwarz1}
\end{align}
where $E$ is an arbitrary real constant. In particular, $W_M$ has zero circulation around each circle $C_j$. The boundary data must furthermore satisfy the compatibility condition
\begin{align*}
\ointctrclockwise_{\partial D_\zeta} T_j\left(\zeta(s),\overline{\zeta(s)}\right) \frac{\partial \sigma_j}{\partial n}\, \d s =0, \qquad  j = 1, \cdots, M,
\end{align*}
where $s$ represents the arc length, $\partial /\partial n$ represents the normal derivative, and $\sigma_j$ represent the \textit{harmonic measures} \citep{Crowdy2008,Crowdy2016a}.

The integral in \eqref{Eq:schwarz1} is taken over every boundary circle and can therefore be decomposed into contributions from each circle. Consequently, the solution to the modified Schwarz problem may also be expressed as
\begin{align*}
W_M(\zeta) &=  -  \frac{1}{2 \pi } \sum_{j=1}^M \ointctrclockwise_{C_j} T_j\left(\zeta^\prime, \overline{\zeta^\prime}\right) \left( \d\log(\omega\left(\zeta^\prime,\zeta \right)) + \d \log\left( \overline{\omega} \left(\overline{\theta_j} \left(\left. 1 \middle/ \zeta \right. \right), \left. 1 \middle/ \zeta^\prime \right. \right)\right)\right)\\
& \frac{1}{2 \pi } \ointctrclockwise_{C_0} M_0\left(\zeta^\prime, \overline{\zeta^\prime}\right) \left( \d\log(\omega\left(\zeta^\prime,\zeta \right)) + \d \log\left( \overline{\omega} \left(\overline{\zeta^\prime}, \left. 1 \middle/ \zeta \right. \right)\right)\right) + E,
\end{align*}
where $\theta_j$ is the M\"obius mapping associated with the $j$-th boundary circle as defined in \eqref{Eq:mobMap}. In the case $M=0$ there is a single object per period window and the solution to the modified Schwarz problem is given by the Poisson formula \citep{Fokas2003}:
\begin{align*}
W_M(\zeta) & = \frac{1}{2 \pi } \ointctrclockwise_{C_0}  M_0\left(\zeta^\prime, \overline{\zeta^\prime}\right) \cdot \frac{\zeta^\prime + \zeta}{\zeta^\prime - \zeta} \cdot  \frac{\d \zeta^\prime}{\zeta^\prime} + E.
\end{align*}
Furthermore, in the case $M=1$ there are two objects per period window, and, when the canonical circular domain is the annulus \mbox{$\rho< \left| z\right|<1$}, the solution is given by the Villat formula \citep{Akhiezer1990,Crowdy2008}
\begin{align*}
W_M(\zeta) & = \frac{1}{2 \pi } \ointctrclockwise_{C_0}  M_0\left(\zeta^\prime, \overline{\zeta^\prime}\right) \cdot \left(1 - 2 K\left(\frac{\zeta^\prime}{\zeta} \right)\right) \cdot \frac{\d \zeta^\prime}{\zeta^\prime}\\
&+ \frac{1}{2 \pi } \ointctrclockwise_{C_1}  M_1 \left(\zeta^\prime, \overline{\zeta^\prime}\right) \cdot 2 K\left(\frac{\zeta^\prime}{\zeta} \right) \cdot  \frac{\d \zeta^\prime}{\zeta^\prime} + E,
\end{align*}
where
\begin{align*}
K(\zeta) &= \frac{\zeta P_\zeta(\zeta)}{P(\zeta)}.
\end{align*}

\subsection{Complex potential for uniform flows, $W_U$}
We now present our solutions for uniform flows through periodic domains. At this point our solutions diverge from those of \cite{Crowdy2010}, and an alternative approach must be taken. In particular, each type of periodic geometries must be considered individually.

\subsubsection{Uniform flows in type I geometries} \label{Sec:uni1}
We now consider the case of uniform flow through a type I geometry.  In the physical plane, the complex potential $w_U$ for uniform flow of strength $U$ inclined at an angle of $\phi$ to the horizontal satisfies
{
	\setlength{\jot}{\baselineskip}
	\begin{align}
	\addtocounter{equation}{1}
	\nabla^2_z w_U(z) & = 0 , & z \in D_z,\tag{\theequation .a}\label{Eq:cZa}\\
	\Im \left[w_{U}(z) \right] &= c_j,  & z \in  L_j, \;\; j = 0, \cdots, M, \tag{\theequation .b}\label{Eq:cZb}\\
	\frac{\d w_U}{\d z} &\sim U \e^{- \i \phi}, & \textrm{as } z  \rightarrow \pm \i \infty, \tag{\theequation .c}\label{Eq:cZc}
	\end{align}
}
where $c_j$ are some constants. We write the complex potential in the $\zeta$-plane as
\begin{align}
w_U(z) & = w_U(f(\zeta)) = W_U(\zeta). \label{Eq:flim}
\end{align}
Consequently, in the $\zeta$-plane, equations \eqref{Eq:cZa}, \eqref{Eq:cZb}, and \eqref{Eq:cZc} become
{
	\setlength{\jot}{\baselineskip}
	\begin{align}
	\addtocounter{equation}{1}
	\nabla^2_\zeta W_U(\zeta) & = 0 , & \zeta \in D_\zeta/\left\{a_{\infty^\pm}\right\}, \tag{\theequation .a}\label{Eq:cZetaa}\\
	\Im \left[W_{U}(\zeta) \right] &= c_j,  & \zeta \in  C_j, \;\; j = 0, \cdots, M, \tag{\theequation .b}\label{Eq:cZetab}\\
	\left. \frac{\d W_U}{\d \zeta}\middle/ \frac{\d f}{\d \zeta } \right.&\sim U \e^{-\i \phi}, & \textrm{as } \zeta \rightarrow a_{\infty^\pm}.\tag{\theequation .c}\label{Eq:cZetac}
	\end{align}
}
Substitution of $\eqref{Eq:typeI}$ into \eqref{Eq:cZetac} yields
\begin{align}
\frac{\d W_U}{\d \zeta}&\sim \pm \frac{U\period\e^{-\i \phi}}{2 \pi \i(\zeta - a_{\infty^\pm})}, \label{Eq:asym}
\end{align}
as $\zeta \rightarrow a_{\infty^{\pm}}$. Consequently, $W_U$ must possess logarithmic branch points at $\zeta = a_{\infty^\pm}$ such that 
\begin{align}
W_{U}(\zeta) \sim \pm \frac{U \period}{2 \pi \i} \e^{- \i \phi} \log(\zeta - a_{\infty^\pm}), \label{Eq:uniAsymp}
\end{align}
in addition to being analytic \eqref{Eq:cZetaa} and taking constant imaginary values on each boundary \eqref{Eq:cZetac}. Using an approach analogous to that of section 6.2 of \cite{Crowdy2010}, we can construct the complex potential by taking linear combinations of point vortices \eqref{Eq:vort2} and source-sink pairs \eqref{Eq:ssPair} located at the preimages of infinity. Note that the complex potentials for point vortices and source-sink pairs are analytic and take constant boundary values on each circle $C_j$. Therefore, the complex potentials each satisfy \eqref{Eq:cZetaa} and \eqref{Eq:cZetab}. We must now strategically choose combinations of source-sink pairs and point vortices to achieve the correct asymptotic behaviour \eqref{Eq:cZetac}. In particular, a unit strength source-sink pair has the asymptotic behaviour
\begin{align}
W_S(\zeta;a_{\infty^+},a_{\infty^-}) \sim \pm \frac{1}{2 \pi } \log(\zeta - a_{\infty^\pm}), \label{Eq:souAsymp}
\end{align}
as $\zeta \rightarrow a_{\infty^\pm}$, whereas a unit strength point vortex has the asymptotic behaviour
\begin{align}
W_V(\zeta;a_{\infty^\pm}) \sim \frac{1}{2 \pi \i} \log \left(\zeta - a_{\infty^\pm}\right). \label{Eq:vortAsymp}
\end{align}
as $\zeta \rightarrow a_{\infty^\pm}$. Comparison of \eqref{Eq:souAsymp} and \eqref{Eq:vortAsymp} with \eqref{Eq:uniAsymp} shows that we may write $W_U$ as
\begin{align*}
W_U &= U \period\left(-\sin(\phi) W_{S;a_{\infty^+}, a_{\infty^-}}(\zeta) + \cos(\phi) \left[W_V(\zeta; a_{\infty^+})  + W_V(\zeta; a_{\infty^-})  \right] \right).
\end{align*}
Note that the no-flux condition \eqref{Eq:cZetab} remains satisfied since we have taken real, linear combinations of the complex potentials $W_S$ and $W_V$. In terms of the prime function, we may expand $W_U$ into the form
\begin{align}
W_U(\zeta) & = \frac{\period U}{2 \pi \i}\left(\e^{- \i \phi} \log\left(\frac{\omega(\zeta, a_{\infty^-}) }{\omega(\zeta, a_{\infty^+})} \right) - \e^{\i \phi} \log\left(\frac{\omega(\zeta, 1/\overline{a_{\infty^-}})}{ \omega(\zeta,1/\overline{a_{\infty^+}})} \right) \right). \label{Eq:sourceSol}
\end{align}
The solution \eqref{Eq:sourceSol} is equivalent to that of appropriately tuned vortex spirals \eqref{Eq:spiral} located at $a_{\infty^\pm}$. It may be shown by analytic continuation and Liouville's theorem that \eqref{Eq:sourceSol} is unique in satisfying \eqref{Eq:cZetaa}, \eqref{Eq:cZetab}, and \eqref{Eq:cZetac}. Moreover, $W_U(\zeta)$ is single-valued as $\zeta$ loops around any of the circles $C_j$, which implies that each object has zero circulation \citep{Crowdy2006b}. Alternatively, \eqref{Eq:sourceSol} can be interpreted as a special form of the vortex spiral introduced in \eqref{Eq:spiral}.

In the case where there is a single object per period window, we have $M=0$ and \eqref{Eq:sourceSol} becomes
\begin{align}
W_U(\zeta) & = \frac{\period U}{2 \pi \i} \left(\e^{- \i \phi} \log\left(\frac{\zeta- a_{\infty^+} }{\zeta-a_{\infty^-}} \right) - \e^{\i \phi} \log\left(\frac{\overline{a_{\infty^+}}}{\overline{a_{\infty^-}}} \cdot \frac{\overline{a_{\infty^-}}\zeta- 1}{ \overline{a_{\infty^+}}\zeta-1}\right) \right). \label{Eq:sourceSolSimp}
\end{align}
\subsubsection*{Comparison to solution for non-periodic flows}
We now show that the solution for uniform flow through a periodic domain \eqref{Eq:sourceSol} collapses to the solution for uniform flow through a non-periodic domain \citep{Crowdy2010} in the large period limit. We take the limit $a_{\infty^+} \rightarrow a_{\infty^-}$, which corresponds to the two branch points in the conformal mapping coalescing to form a simple pole. Consequently, in this limit there is no branch cut and the mapping is not periodic. Taylor expanding the logarithms in \eqref{Eq:sourceSol} yields 
\begin{align}
\addtocounter{equation}{1}
\log\left(\frac{\omega(\zeta, a_{\infty^+}) }{\omega(\zeta, a_{\infty^-})} \right) \sim  \left( a_{\infty^-} - a_{\infty^+} \right) &\frac{\omega_{\phi}(\zeta, a_{\infty^-})}{\omega(\zeta, a_{\infty^-})} \notag\\
&= 2 \pi \i  \left( a_{\infty^-} - a_{\infty^+} \right) \left[ \frac{\partial G_0}{\partial \alpha}(\zeta, \alpha) \right]_{\alpha = a_{\infty^-}} ,\label{Eq:lim1} \tag{\theequation .a} \\
\log\left(\frac{\omega(\zeta, 1/\overline{a_{\infty^-}})}{ \omega(\zeta,1/\overline{a_{\infty^+}})} \right) \sim  \left( a_{\infty^-} - a_{\infty^+} \right) &\frac{1}{\overline{a_{\infty^-}}^2}\frac{\omega_{\alpha}(\zeta, 1/\overline{a_{\infty^-}})}{\omega(\zeta, 1/\overline{a_{\infty^-}})} \notag\\
& = 2 \pi \i  \left( a_{\infty^-} - a_{\infty^+} \right) \left[ \frac{\partial G_0}{\partial \overline{\alpha}}( \zeta ,\alpha) \right]_{\alpha = a_{\infty^-}}, \label{Eq:lim2} \tag{\theequation .b}
\end{align}
subject to addition of an arbitrary constant, where the subscript $\alpha$ indicates the derivative with respect to the second argument. Substitution of  \eqref{Eq:lim1} and \eqref{Eq:lim2} into \eqref{Eq:sourceSol} and rescaling $\period =  \tilde{\period}/ \left(a_{\infty^+}  - a_{\infty^-} \right) $ yields
\begin{align*}
W_U(\zeta) \sim 2 \pi \i \tilde{\period}U  \left[ \e^{\i \phi} \frac{\partial G_0}{\partial \overline{\alpha}} - \e^{ -\i \phi}  \frac{\partial G_0}{\partial {\alpha}} \right]_{\alpha = a_{\infty^-}},
\end{align*}
as $a_{\infty^+} \rightarrow a_{\infty^-}$, which is equivalent to the complex potential for uniform flow through a non-periodic system as detailed in section 6.2 of \cite{Crowdy2010}.

\subsubsection{Uniform flows in type II geometries} \label{Sec:uni2}

The uniform flow for a type II geometry may be calculated in a similar fashion to that in section \ref{Sec:uni1} by strategically placing flow singularities in the circular domain. In a type II geometry, the flow angle in the far field is restricted to be parallel to the period. Therefore, the conditions on $w_U$ are equivalent to \eqref{Eq:cZa} and \eqref{Eq:cZb}, but \eqref{Eq:cZc} is replaced with
\begin{align*}
\frac{\d w_U}{\d z} \sim U, \qquad \qquad \qquad \textrm{as } z \rightarrow +\i \infty,
\end{align*}
since we assume the period to be real and positive.
Accordingly, instead of placing a vortex spiral \eqref{Eq:spiral} at the preimage of infinity, we now place a point vortex \eqref{Eq:G0} at the preimage of infinity and may therefore write the solution as
\begin{align}
W_U(\zeta) &= \frac{U\period}{2 \pi \i} \log \left(\frac{\omega(\zeta,a_\infty)}{\left| a_\infty\right| \omega(\zeta, 1/\overline{a_\infty})}\right).
\end{align}
Note that this solution induces circulation $U\period$ on the circle $C_0$, which corresponds to the lower boundary of the target domain. 
\subsubsection{Uniform flows in type III geometries} \label{Sec:uni3}
Constructing the potential for a uniform flow in a case III geometry requires a different approach to that of type I and type II since the period window is bounded. Instead we construct a flow that has a specified circulation on the boundaries $L_0$ and $L_1$ of the period window. Additionally, since no point is mapped to infinity, there can be no singularities of the complex potential in the circular domain. We may express the complex potential for such a flow in the form
\begin{align}
W_U(\zeta) &= U \period(G_1(\zeta,\alpha) - G_0(\zeta,\alpha))\notag \\
&= \frac{U \period}{2 \pi \i} \log\left(\frac{\omega(\zeta,1/\overline{\alpha})}{\omega(\zeta,\theta_1(1/\overline{\alpha}))}\right), \label{Eq:uni3}
\end{align}
for any $\alpha \in D_\zeta$, and where $U \period$ is now the circulation on $C_0$ and $C_1$. Note that these quantities are related to the \emph{harmonic differentials} $v_j$ (also referred to as the \emph{first kind integrals}) by lemma 5.1 of \cite{Crowdy2016a}, which gives
\begin{align*}
G_1(\zeta, \alpha) - G_0(\zeta,\alpha) &= - v_j(\zeta) + \overline{v_j(\alpha)} +\frac{1}{2} \tau_{jj} + \frac{1}{2 \pi} \arg \left[ \frac{\alpha}{\alpha - \delta_1}\right],
\end{align*}
where $\tau_{jj}$ is a constant.

The flux through a single period window is related to the circulation on $C_0$ and $C_1$. Whilst the circulation is given by the jump in velocity potential on, for example, $C_0$,
\begin{align*}
U = \bigg[\Re \left[W_U \right]\bigg]_{\zeta \in C_0},
\end{align*}
the flux, $Q$, is given by the jump in streamfunction between $C_0$ and $C_1$:
\begin{align*}
Q = \bigg[\Im \left[W_U\right]\bigg]_{\zeta \in C_0}^{\zeta \in C_1}.
\end{align*}
In general, there are not closed form relationships between $U$ and $Q$. However, in the doubly connected case for a concentric annulus of interior radius $q$, \eqref{Eq:uni3} becomes 
\begin{align*}
W_U(\zeta) = \frac{U\period}{2 \pi \i} \log(\zeta),
\end{align*}
and we may relate the flux to the circulation as
\begin{align*}
Q = \frac{U\period}{2 \pi} \log(q).
\end{align*}

\subsection{Straining flows}
Our calculus may also consider straining flows where the flow velocity tends to infinity in the far-field.
\subsubsection{Straining flows in type I geometries}
We consider a straining flow whose complex potential has the far-field behaviour
\begin{align}
w_S^\pm(z) \sim S_\pm \e^{-\i \chi_\pm} \e^{\mp 2 \pi \i z/\period}, \label{Eq:strain1}
\end{align}
as $z \rightarrow \pm \i \infty$ for real constants $S_\pm$, $\chi_\pm$. In the circular domain, \eqref{Eq:strain1} becomes
\begin{align*}
W_S^\pm(\zeta) \sim \frac{S_{\pm} \e^{-\i \chi_\pm}}{\zeta - a_{\infty^\pm}}.
\end{align*}
By comparison with \eqref{Eq:doubAsymp}, the  complex potentials $W_S^\pm$ may be constructed by taking appropriately placed doublets and writing
\begin{align}
W_S^\pm (\zeta) &= S_{\pm} W_D(\zeta; \chi_\pm) \notag \\
& = 2 \pi \i  S_{\pm} \left[\e^{\i \lambda} \frac{\partial G_0}{\partial \overline{\alpha}}- \e^{-\i \lambda} \frac{\partial G_0}{\partial {\alpha}}\right]_{\alpha = a_{\infty^\pm}}. \label{Eq:strainSol1}
\end{align}
In an analogous way to section \ref{Sec:uni1}, it may be shown that we recover the original straining flow solution by \cite{Crowdy2010} by taking the limit $a_{\infty^+} \rightarrow a_{\infty^-}$. Note that the solution \eqref{Eq:strainSol1} induces zero circulation around every boundary.
 \subsubsection{Straining flows in type II geometries}
The complex potential for a straining flow in a type II geometry has the asymptotic behaviour
 \begin{align}
 w_S(z) \sim S \e^{-\i \chi} \e^{-2 \pi \i z/\period}. \label{Eq:strain1}
 \end{align}
 as $z \rightarrow + \i \infty$ for real constants $S$, $\chi$. The solution takes essentially the same form as that of \eqref{Eq:strainSol1}:
\begin{align*}
W_S (\zeta) &= 2 \pi \i  S  \left[\e^{\i \lambda} \frac{\partial G_0}{\partial \overline{\alpha}}- \e^{-\i \lambda} \frac{\partial G_0}{\partial {\alpha}}\right]_{\alpha = a_{\infty}}.
\end{align*}
 \subsubsection{Straining flows in type III geometries}
We do not consider straining flows in type III geometries since the domain is bounded.
\subsection{Complex potential for circulatory flows, $W_\Gamma$}
Thus far, we have deliberately only constructed solutions with vanishing circulation around each object. The exception was the solution for point vortices introduced in section \ref{Sec:vort}, whose strengths affected the circulation around the boundaries. We now explain how to tune the circulation around each object in the absence of singularities in the flow.
\subsubsection{Circulatory flows in type I geometries} \label{Sec:circ1}
In case I geometries, we may modify the circulation around each boundary by strategically placing vortices at the preimages of infinity. To change the circulation around $C_j$, we use the modified hydrodynamic Green's functions which were introduced in \eqref{Eq:modGreen}. The complex potential for a circulatory flow in a periodic domain with no singularities in the flow, is given by
\begin{align*}
W_\Gamma(\zeta) & = -\Gamma_0^+ G_0(\zeta, a_{\infty^+}) -\Gamma_0^- G_0(\zeta, a_{\infty^-})\\
&+ \sum_{j=1}^M \Gamma_j^- G_j(\zeta,a_{\infty^+}) +\Gamma_j^+ G_j(\zeta,a_{\infty^-})
\end{align*}
for $j = 0,\cdots, M$. The potential $W_\Gamma$ induces circulation $\Gamma_j = \Gamma_j^+ + \Gamma_j^-$ around each boundary $L_j$. The reason for splitting $\Gamma_j$ in to these two parts is that by placing vortices at $a_{\infty^\pm}$ we are changing the behaviour in the far field. In particular, the far-field behaviour is now
\begin{align*}
\frac{\d w_\Gamma}{\d z} \sim \pm \Gamma^\pm \period \qquad \qquad \textrm{as } z \rightarrow  \pm \i\infty .
\end{align*}
where $\Gamma^\pm = \sum_{j=0}^M \Gamma_j^\pm$.

\subsubsection{Circulatory flows in type II geometries}
We employ a similar strategy to the previous section to modify the circulation around the boundaries in type II geometries. In section \ref{Sec:uni2} we showed that a uniform flow may be constructed by placing a point vortex at the preimage of infinity. This point vortex induced a circulation around the boundary $L_0$, which represented the lower boundary of the domain. To specify the circulations around the boundaries $L_j$ for $j = 1, \cdots M,$ we instead use the modified hydrodynamic Green's functions and write
\begin{align*}
W_\Gamma(\zeta) & = \sum_{j=1}^M \Gamma_j G_j(\zeta,a_{\infty}).
\end{align*}
This complex potential induces circulation $\Gamma_j$ around the boundary $L_j$. Note that this complex potential affects the flow in the far field, which is now
\begin{align*}
\frac{\d w_\Gamma}{\d z} \sim \Gamma \period \qquad \qquad \textrm{as } z \rightarrow  +\i \infty.
\end{align*}
where $\Gamma = \sum_{j = 1}^M \Gamma_j$
\subsubsection{Circulatory flows in type III geometries}
In section \ref{Sec:uni3} we showed that uniform flows through type III geometries may be obtained by specifying an equal circulation on boundaries $L_0$ and $L_1$. We now show how to specify the circulation around interior boundaries $L_j$ for $j>2$. Since the domain is bounded, the net circulation around the interior boundaries must vanish. Otherwise, integrating the complex velocity around the boundaries would result in a non-zero quantity, thereby violating Cauchy's theorem as there are no singularities in the flow. In type 3 geometries, the circulation around interior boundaries may again be specified with the modified hydrodynamics Green's function. For example, the complex potential with circulation $\Gamma_j$ around $L_j$ for $j>2$ is given by
\begin{align*}
W_\Gamma(\zeta) & = \sum_{j=2}^M \Gamma_j G_j(\zeta, \gamma),
\end{align*}
for any $\gamma \in D_\zeta$. The net circulation around the interior boundaries must vanish, so $\sum_{j=2}^M \Gamma_j = 0$.

\subsubsection{The Kutta condition}
In many applications, certain solutions produced by potential flow analyses are undesirable. In aerodynamics, the flow at the trailing edge of a body must be finite \citep{Crighton1985,Eldredge2019a}, and satisfying this requirement is known as the \emph{Kutta condition}. We can express the total complex potential as 
\begin{align*}
W(\zeta) = W_\Gamma(\zeta) + \tilde{W}(\zeta),
\end{align*}
where $\tilde{W}$ represents the complex potential with vanishing circulation. Enforcing the Kutta condition requires the velocity at each trailing edge to vanish. If the trailing edge of the boundary $L_j$ is located at $\zeta_j$ in the circular domain, the Kutta condition requires
\begin{align}
\frac{\d W_\Gamma}{\d \zeta}(\zeta_j) + \frac{\d \tilde{W}}{\d \zeta}(\zeta_j) = 0, \qquad \qquad j = 0, \cdots M. \label{Eq:Kutta}
\end{align}
Since $W_\Gamma$ contains $M+1$ unknown circulations, \eqref{Eq:Kutta} represents an $(M+1)\times(M+1)$ linear system of equations that can easily be inverted.

As noted in section \ref{Sec:circ1}, in case I geometries, the effect of modifying the circulations is to modify the angle and strength of the flow far away from the objects. Similarly, the effect of the Kutta condition is to modify the far-field behaviour of the flow. For example, if an inlet flow angle is specified, the Kutta condition is used to determine the outlet angle.

\section{Conclusion} \label{Sec:conclusions}
In this paper we have presented a constructive procedure for calculating 2-D potential flows through periodic domains of arbitrary geometry and connectivity. The first step in the procedure is to construct a conformal mapping from a circular domain to the physical target domain. A constructive formula for such mappings is available in \cite{Baddoo2019c}. The second step in the procedure is to construct the complex potential for the flow in the circular domain. By constructing both  the conformal mapping and the complex potential in terms of the Schottky--Klein prime function, the ensuing solutions are valid for arbitrary connectivities. An essential feature of the modelling process is the correct identification of physically relevant potential flow phenomena. Although the calculus we have presented accounts for a range of flow types, of particular note are the complex potentials for uniform flow and point vortices: the former can be used to model the motion of the periodic array through a fluid (or vice-versa), whilst the latter can be used to represent discrete quantities of vorticity.

The solutions we have found in this paper could be applied to a range of physical scenarios. In particular, potential flows may be used to accurately model unsteady inviscid flows in aerospace applications \citep{Eldredge2019a}. Since the potential flow solutions struggle to  account for viscosity, researchers are now turning to data-driven methods to incorporate missing physics. Whilst it is generally accepted that the Kutta condition is appropriate to constrain the flow at the trailing edge, there is some debate regarding the flow condition at the leading edge. After a detailed numerical and experimental campaign, \cite{Ramesh2014} proposed a leading-edge suction parameter criterion to control the amount of vorticity shed from the leading edge. This criterion was then applied by \cite{Darakananda2018a} to couple a reduced-order model \citep{Darakananda2019} with high-fidelity ``truth'' simulation \citep{Taira2007} to calculate the lift on a flat plate undergoing a sudden motion. Such a data-driven approach could certainly be applied to the geometries in the present work to model flow separation in turbomachinery stages. This will be a topic of future work.

\begin{acknowledgements}
	P. J. B. acknowledges support from EPSRC grant 1625902. L. J. A. acknowledges support from EPSRC grant EP/P015980/1.
\end{acknowledgements}

\section*{Conflict of interest}
The authors declare that they have no conflict of interest.

\appendix
\section{Rigid body motion kinematic condition} \label{Ap:rigid}
In this section we derive the kinematic boundary condition for the complex potential due to a rigid body motion. In the physical plane, the complex potential induced by the motion of the boundaries is given by $w_M$. Following \cite{Crowdy2008a}, we now manipulate \eqref{Eq:KinBCVec} to derive a condition on the imaginary part of $w_M$ on the boundaries. We note that the dot product may be rewritten in terms of complex notation as $\mathbf{a}\cdot \mathbf{b} = \Re[ \bar{a} b]$ where $a$ is the obvious complexification of $\mathbf{a}$. We also note that the tangent vector $\mathbf{t}$ may be written in complex form as $\d z /\d s$, where $s$ is the arc length. Therefore, the normal vector $\mathbf{n}$ may be written as $-\i \d z / \d s$ and the kinematic boundary condition \eqref{Eq:KinBCVec} may be written as 
\begin{align*}
\Re\left[\bar{u}(s) \times -\i \frac{\d z}{\d s}\right] &= \Re \left[ \bar{U}_j(s)\times -\i \frac{\d z}{\d s}\right].
\end{align*}
Using the standard representation of complex velocity, this expression becomes
\begin{align}
\Re\left[\frac{\d w_M}{\d z} \times -\i \frac{\d z}{\d s}\right] &= \Re \left[ \bar{U}_j(s)\times -\i \frac{\d z}{\d s}\right]. \label{Eq:kin1}
\end{align}
The first term may be simplified by the chain rule to obtain
\begin{align*}
\Re\left[ -\i \frac{\d w_M}{\d s} \right] &= \Re \left[-\i \bar{U}_j(s) \frac{\d z}{\d s}\right].
\end{align*}
In the present work, we only consider rigid bodies. Accordingly, the only possible motions are rotations and translations. Therefore, every point $z\in  P_j$ may be expressed as
\begin{align*}
z = c_j(t) + \eta_j(s)\e^{\i \theta_j(t)}.
\end{align*}
The velocity of each moving object may therefore be written as
\begin{align*}
U_j(s) &= \dot{c}_j(t) + \i \dot{\theta}_j(t) \eta_j(s) \e^{\i \theta_j(t)} =  \dot{c}_j(t) + \i \dot{\theta}_j(t) \left(z-c_j(t)\right),
\end{align*}
Therefore, the kinematic condition \eqref{Eq:kin1} becomes
\begin{align}
\Re\left[ -\i \frac{\d w_M}{\d s} \right] &= 
 \Re \left[-\i \dot{\bar{c}}_j(t)\frac{\d z}{\d s} - \dot{\theta}_j(t) \left(\bar{z}-\bar{c}_j(t)\right) \frac{\d z}{\d s}\right]. \label{Eq:kin2}
\end{align}
Noting that 
\begin{align*}
\frac{\d }{\d s} \left| z-c_j\right|^2 &= 2 \Re\left[\frac{\d z}{\d s} \left(\bar{z}- \bar{c}_j\right)\right],
\end{align*}
we may integrate the kinematic condition  \eqref{Eq:kin2} with respect to arc length $s$ to get
\begin{align}
\Re\left[ -\i w_M \right] &= \Re \left[-\i \dot{\bar{c}}_j(z-c_j) - \frac{\dot{\theta}_j}{2} \left| z-c_j\right|^2 \right] + d_j,  \qquad \textnormal{for } z \in  L_j \label{Eq:IntKinBC}
\end{align}
for a constant $d_j$ are chosen to comply with a compatibility condition \citep{Crowdy2008}.

where $s$ represents the arc length, $\partial /\partial n$ represents the normal derivative, and $\sigma_j$ represent the \textit{harmonic measures} \citep{Crowdy2008,Crowdy2016a}.

We write $W_M(\zeta)=w_M(z(\zeta))$ and translate \eqref{Eq:IntKinBC} into the canonical circular domain to obtain the condition
\begin{align}
\Im\left[ W_M(\zeta) \right] &= \Im \left[ \dot{\bar{c}}_j(z(\zeta)-c_j) - \i \frac{\dot{\theta}_j}{2} \left| z(\zeta)-c_j\right|^2 \right]+d_j \equiv T_j \left(\zeta,\bar{\zeta} \right), \qquad \textnormal{for } \zeta \in C_j. \label{Eq:IntKinBCZeta}
\end{align}
Finally, the boundary data must furthermore satisfy the compatibility conditions
\begin{align*}
\ointctrclockwise_{\partial D_\zeta} T_j\left(\zeta(s),\overline{\zeta(s)}\right) \frac{\partial \sigma_j}{\partial n}\, \d s =0, \qquad  j = 1, \cdots, M.
\end{align*}

\def\myPathTextAbove#1#2#3#4{
	\draw [-LaTeX,thick,postaction={decorate,
		decoration={
			raise=2ex,
			text along path,
			text align={center},
			text={%
				|\color{black}| {#1} }
		}
	}
	] #2 to [bend left=#4] #3;
}

\section{Catalogue of periodic conformal slit maps} 
In this section we present a range of new conformal slit maps from circular domains for periodic domains. The definitions presented herein are exact and do not make use of any approximations. Moreover, the mappings are valid for arbitrary connectivities and all types of periodic map: types I, II and III. The basic approach in this section is to construct a mapping that relates a multiply connected circular domain to an intermediate slit domain made up of either circular arc slits, radial slits, or both. The circular domain is labelled $D_\zeta$, the intermediate domain $D_\xi$, and the target domain $D_z$. Taking a logarithm of the $\xi$-plane then results in periodic slit mapping in the $z$-plane.

\subsection{Type I periodic conformal slit maps} \label{Ap:conf1}

\subsubsection{Type I parallel slit map} \label{Sec:confIa}
\begin{figure}[!h]
	\def\stag{3}
	\centering
	\begin{tikzpicture}[scale = .7]

	\begin{scope}[shift = {(-.7\linewidth,0)},scale =.5]
	
	\path [path fading = circle with fuzzy edge 15 percent, fill=confgrey, even odd rule] (0,0) circle (\zeroRad) (0,0) circle (1.3*\zeroRad);
	\draw[blue,fill= confcol,line width = \lw] (0,0) circle (\zeroRad);
	\node at (-.5,1.85) {$D_\zeta$};
	
	\skCircBlankBThick{(\delIx,\delIy)}{\radI}{red}
	\skCircBlankBThick{(\delIIx,\delIIy)}{\radII}{green}
	\skCircBlankBThick{(\delIIIx,\delIIIy)}{\radIII}{magenta}
	
		\fill[branchCol] (1.5,-.5) circle (.2);
				\fill[branchCol] (-1.2,-1) circle (.2);
	
	\end{scope}
	
	\begin{scope}[shift = {(.35\linewidth,0)}, rotate = 90, xscale=.4, yscale = .7]

	\foreach \x in {-1,0,1}
	{
		\draw[dashed,thick,branchCol] (-5,1+\stag*\x)--(5,1+\stag*\x);
	}
	
	\fill[white, path fading = north] (-5,-3) rectangle (-3,6);
	
	\fill[white, path fading = south] (3,6) rectangle (5,-3);
	\node at (1.8,2.5) {$D_z$};
	
	\foreach \x in {-1,0,1,2}
	{
		\draw[green,line width = 2*\lw] (-1,-.8+\stag*\x)--(-1,.7+\stag*\x);
		\draw[blue,line width = 2*\lw] (-2.5,-1.5+\stag*\x)--(-2.5,.5+\stag*\x);
				\draw[magenta,line width = 2*\lw] (2.8,.7+\stag*\x)--(2.8,-1.4+\stag*\x);
				\draw[red,line width = 2*\lw] (0,.5+\stag*\x)--(0,-.9+\stag*\x);
	}
	
	\fill[white,path fading=  west] (-5,-\stag-.01) rectangle (5,-\stag/2);
	\fill[white] (-5,-\stag/2-\stag) rectangle (5,-\stag);
	
	\fill[white,path fading = east] (-5,2*\stag+0.01) rectangle (5,\stag + \stag/2);
	\fill[white] (-5,\stag/2+2*\stag) rectangle (5,2*\stag);

	\end{scope}

	\begin{scope}[shift = {(-.3\linewidth,0)}, scale =.5]

				\draw[branchCol,dashed,line width = \lw] (0,0)--(\zeroRad,0);
	\fill[branchCol] (0,0) circle (.2);
	\draw [green,domain=160:330,line width = 2*\lw] plot ({2*cos(\x)}, {2*sin(\x)});
		\draw [blue,domain=120:250,line width = 2*\lw] plot ({3*cos(\x)}, {3*sin(\x)});
		\draw [magenta,domain=30:100,line width = 2*\lw] plot ({1.4*cos(\x)}, {1.4*sin(\x)});
			\draw [red,domain=120:200,line width = 2*\lw] plot ({1.6*cos(\x)}, {1.6*sin(\x)});
	\node at (.8,2) {$D_\xi$};

	\end{scope}

		\myPathTextAbove{$\xi = A_{\rm I}(\zeta;a_{\infty^+},a_{\infty^-})$}{(-6.75,1.7)}{(-5.25,1.7)}{20}
	
			\myPathTextAbove{$z = \period \log(\xi)/(2\i\pi)$}{(-1.75,1.7)}{(-.25,1.7)}{20}
			\node at (0,2.5) {};
	\end{tikzpicture}
	\caption{A type I periodic parallel slit map for $M=3$.}
	\label{Fig:confIa}
\end{figure}
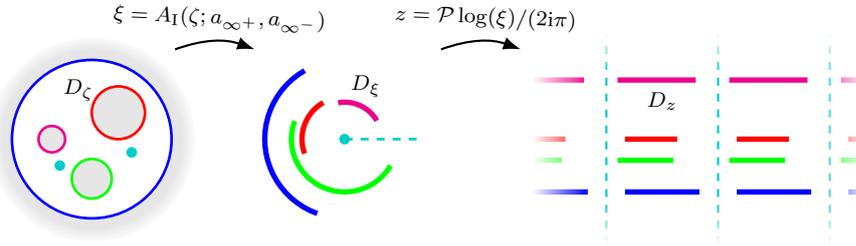

We first introduce the type I circular arc slit map, $A_{\rm I}$, defined in section 7.2 of \cite{Crowdy2006} as
\begin{align}
\xi = A_{\rm I}(\zeta;a_{\infty^+},a_{\infty^-}) \equiv \frac{\omega(\zeta,a_{\infty^+})\omega(\zeta,1/\overline{a_{\infty^-}})}{\omega(\zeta, a_{\infty^-})\omega(\zeta,1/\overline{a_{\infty^+}})}. \label{Eq:circularSlitI}
\end{align}
This mapping relates each boundary circle to a circular slit of finite length, as illustrated in figure \ref{Fig:confIa}. The point $a_{\infty^+}$ is mapped to the origin, and the point $a_{\infty^-}$ is mapped to infinity. The angle and length of the slits depends on the locations of the boundary circles in the preimage circular domain and the choices of $a_{\infty^\pm}$. Accordingly, taking the scaled logarithm
\begin{align}
z = \frac{\period}{2 \pi \i} \log\left(\xi\right) \label{Eq:log}
\end{align}
transplants each circular arc onto a periodically repeated horizontal slit. Again, the mapping is illustrated in figure \ref{Fig:confIa}.

\subsubsection{Type I periodic perpendicular slit map}
\begin{figure}[!h]
	\def\stag{3}
	\centering
	\begin{tikzpicture}[scale = .7]

	\begin{scope}[shift = {(-.7\linewidth,0)},scale =.5]
	
	\path [path fading = circle with fuzzy edge 15 percent, fill=confgrey, even odd rule] (0,0) circle (\zeroRad) (0,0) circle (1.3*\zeroRad);
	\draw[blue,fill= confcol,line width = \lw] (0,0) circle (\zeroRad);
	\node at (-.5,1.85) {$D_\zeta$};
	
	\skCircBlankBThick{(\delIx,\delIy)}{\radI}{red}
	\skCircBlankBThick{(\delIIx,\delIIy)}{\radII}{green}
	\skCircBlankBThick{(\delIIIx,\delIIIy)}{\radIII}{magenta}
	
		\fill[branchCol] (1.5,-.5) circle (.2);
				\fill[branchCol] (-1.2,-1) circle (.2);
	
	\end{scope}
	
	\begin{scope}[shift = {(.35\linewidth,0)}, rotate = 90, xscale=.4, yscale = .7]

	\foreach \x in {-1,0,1}
	{
		\draw[dashed,thick,branchCol] (-5,1+\stag*\x)--(5,1+\stag*\x);
	}
	
	\fill[white, path fading = north] (-5,-3) rectangle (-3,6);
	
	\fill[white, path fading = south] (3,6) rectangle (5,-3);
	\node at (1.8,3) {$D_z$};
	
	\foreach \x in {-1,0,1,2}
	{
		\draw[green,line width = 2*\lw] (-1,0+\stag*\x)--(1,0+\stag*\x);
		\draw[blue,line width = 2*\lw] (-2.5,.5+\stag*\x)--(2.5,.5+\stag*\x);
				\draw[magenta,line width = 2*\lw] (-2.8,-1+\stag*\x)--(.8,-1+\stag*\x);
				\draw[red,line width = 2*\lw] (0,-.5+\stag*\x)--(2,-.5+\stag*\x);
	}
	
	\fill[white,path fading=  west] (-5,-\stag-.01) rectangle (5,-\stag/2);
	\fill[white] (-5,-\stag/2-\stag) rectangle (5,-\stag);
	
	\fill[white,path fading = east] (-5,2*\stag+0.01) rectangle (5,\stag + \stag/2);
	\fill[white] (-5,\stag/2+2*\stag) rectangle (5,2*\stag);

	\end{scope}

	\begin{scope}[shift = {(-.3\linewidth,0)}, scale =.5]

				\draw[branchCol,dashed,line width = \lw] (0,0)--(\zeroRad,0);
	\fill[branchCol] (0,0) circle (.2);
	\draw [blue,domain=.5:3,line width = 2*\lw] plot ({\x*cos(30)}, {\x*sin(30});
		\draw [green,domain=1:3,line width = 2*\lw] plot ({\x*cos(100)}, {\x*sin(100)});
		\draw [red,domain=1.5:3,line width = 2*\lw] plot ({\x*cos(200)}, {\x*sin(200)});
			\draw [magenta,domain=.5:3,line width = 2*\lw] plot ({\x*cos(300)}, {\x*sin(300)});
	\node at (.8,2) {$D_\xi$};

	\end{scope}

		\myPathTextAbove{$\xi = R_{\rm I}(\zeta; a_{\infty^+}, a_{\infty^-})$}{(-6.75,1.7)}{(-5.25,1.7)}{20}
	
			\myPathTextAbove{$z = \period \log(\xi)/(2\i\pi)$}{(-1.75,1.7)}{(-.25,1.7)}{20}
			\node at (0,2.5) {};
	\end{tikzpicture}
	\caption{A typical type I periodic perpendicular slit map $M=3$.}
	\label{Fig:confIb}
\end{figure}
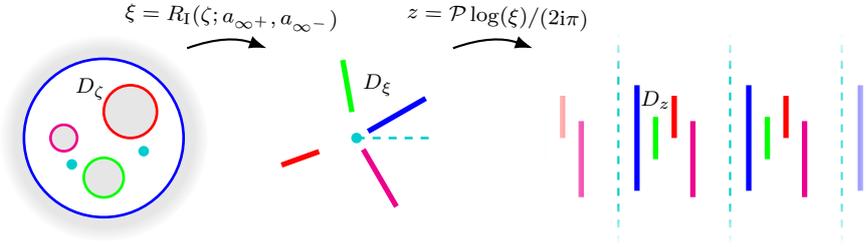

We now introduce the radial slit map defined in section 7.5 of \cite{Crowdy2006}:
\begin{align*}
\xi =R_{\rm I}(\zeta; a_{\infty^+}, a_{\infty^-}) & \equiv \frac{\omega(\zeta, a_{\infty^+})\omega(\zeta,1/\overline{a_{\infty^+}})}{\omega(\zeta,a_{\infty^-})\omega(\zeta,1/\overline{a_{\infty^-}})}.\label{Eq:radialSlit}
\end{align*}
This mapping transplants each boundary circle onto a finite radial slit, as illustrated in figure \ref{Fig:confIb}. Similarly to the circular arc slit map \eqref{Eq:circularSlitI}, the lengths and angles of the radial slits depends on the conformal modulii and $a_{\infty^\pm}$. In an analogous way to section \ref{Sec:confIa}, applying the scaled logarithm \eqref{Eq:log} transplants each radial slit to a periodically repeated vertical slit, as illustrated in figure \ref{Fig:confIb}.
\subsubsection{Type I periodic angled slit map}
\begin{figure}[!h]
	\def\stag{3}
	\centering
	\begin{tikzpicture}[scale = .7]

	\begin{scope}[shift = {(-.5\linewidth,0)},scale =.5]
	
	\path [path fading = circle with fuzzy edge 15 percent, fill=confgrey, even odd rule] (0,0) circle (\zeroRad) (0,0) circle (1.3*\zeroRad);
	\draw[blue,fill= confcol,line width = \lw] (0,0) circle (\zeroRad);
	\node at (-.5,1.85) {$D_\zeta$};
	
	\skCircBlankBThick{(\delIx,\delIy)}{\radI}{red}
	\skCircBlankBThick{(\delIIx,\delIIy)}{\radII}{green}
	\skCircBlankBThick{(\delIIIx,\delIIIy)}{\radIII}{magenta}
	
		\fill[branchCol] (1.5,-.5) circle (.2);
				\fill[branchCol] (-1.2,-1) circle (.2);
	
	\end{scope}
	
	\begin{scope}[shift = {(.3\linewidth,0)}, rotate = 90, xscale=.5, yscale = .7]

	\foreach \x in {-1,0,1}
	{
		\draw[dashed,thick,branchCol] (-5,1+\stag*\x)--(5,1+\stag*\x);
	}
	
	\fill[white, path fading = north] (-5,-3) rectangle (-3,6);
	
	\fill[white, path fading = south] (3,6) rectangle (5,-3);
	\node at (1.8,2.5) {$D_z$};
	
	\foreach \x in {-1,0,1,2}
	{
		\draw[green,line width = 2*\lw, shift= {(0,\stag*\x)},rotate = 30] (-1,-.8)--(-1,.7);
		\draw[blue,line width = 2*\lw, shift = {(0,\stag*\x+1.2)},rotate = 30] (-2.5,-1.5)--(-2.5,.5);
				\draw[magenta,line width = 2*\lw,shift = {(0,\stag*\x-1.2)},rotate = 30] (2.8,.7)--(2.8,-1.4);
				\draw[red,line width = 2*\lw,shift = {(0,\stag*\x)},rotate = 30] (0,.75)--(0,-.9);
	}
	
	\fill[white,path fading=  west] (-5,-\stag-.01) rectangle (5,-\stag/2);
	\fill[white] (-5,-\stag/2-\stag) rectangle (5,-\stag);
	
	\fill[white,path fading = east] (-5,2*\stag+0.01) rectangle (5,\stag + \stag/2);
	\fill[white] (-5,\stag/2+2*\stag) rectangle (5,2*\stag);
	
	\end{scope}

		\myPathTextAbove{$\xi = S(\zeta; a_{\infty^+},a_{\infty^-},\chi,\period)$}{(-3.4,1.5)}{(-1.4,1.5)}{20}
	
			\node at (0,2) {};
	\end{tikzpicture}
	\caption{A typical angled periodic slit map for $M=3$.}
	\label{Fig:confIc}
\end{figure}
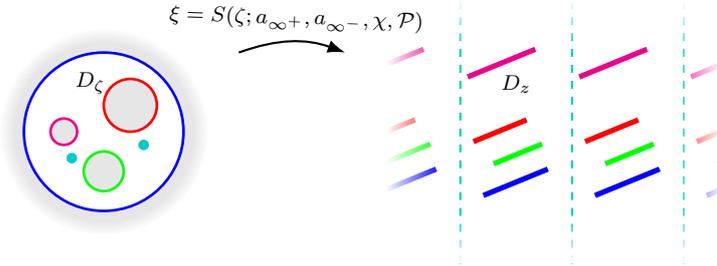

We now take linear combinations of the parallel and perpendicular slit maps to obtain mappings for angled slits. In particular, the mapping
\begin{align}
	\zeta &= \frac{\period}{2 \pi \i} \left[\cos(\chi)  \log\left(A_{\rm I}(\zeta; a_{\infty^+},a_{\infty^-})\right) -\i \sin(\chi) \log\left(R_{\rm I}(\zeta, a_{\infty^+},a_{\infty^-}) \right)  \right]\e^{\i \chi}\notag \\
	&= \frac{\period }{2 \pi \i} \left[\log\left(\frac{\omega(\zeta, a_{\infty^+})}{\omega(\zeta,a_{\infty^-})}\right) -\e^{2\i \chi} \log\left(\frac{\omega(\zeta, 1/\overline{a_{\infty^+}})}{\omega(\zeta,1/\overline{a_{\infty^-}})}\right)  \right]
	\end{align}
maps a circular domain to a periodic array of slits inclined at angle $\chi$ to the horizontal with real period $\period$. Such a mapping is illustrated in figure \ref{Fig:confIc}.

\subsection{Type II periodic conformal slit maps} \label{Ap:conf2}

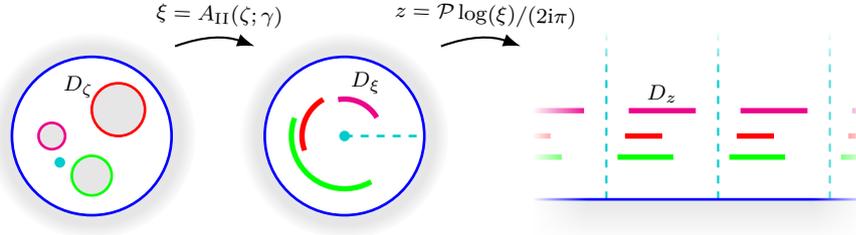
\begin{figure}[!h]
	\def\stag{3}
	\centering
	\begin{tikzpicture}[scale = .7]

	\begin{scope}[shift = {(-.7\linewidth,0)},scale =.5]
	
	\path [path fading = circle with fuzzy edge 15 percent, fill=confgrey, even odd rule] (0,0) circle (\zeroRad) (0,0) circle (1.3*\zeroRad);
	\draw[blue,fill= confcol,line width = \lw] (0,0) circle (\zeroRad);
	\node at (-.5,1.85) {$D_\zeta$};

	\skCircBlankBThick{(\delIx,\delIy)}{\radI}{red}
	\skCircBlankBThick{(\delIIx,\delIIy)}{\radII}{green}
	\skCircBlankBThick{(\delIIIx,\delIIIy)}{\radIII}{magenta}
					\fill[branchCol] (-1.2,-1) circle (.2);
	\end{scope}
	
	\begin{scope}[shift = {(.35\linewidth,0)}, rotate = 90, xscale=.4, yscale = .7]

	\foreach \x in {-1,0,1}
	{
		\draw[dashed,thick,branchCol] (-3,1+\stag*\x)--(5,1+\stag*\x);
	}
	
	\fill[confgrey, path fading = south] (-5,-3) rectangle (-3,6);
	\draw[blue, line width = \lw] (-3,-3)--(-3,6);
	
	\fill[white, path fading = south] (3,6) rectangle (5,-3);
	\node at (2,2.5) {$D_z$};
	
	\foreach \x in {-1,0,1,2}
	{
		\draw[green,line width = 2*\lw] (-1,-.8+\stag*\x)--(-1,.7+\stag*\x);
		\draw[magenta,line width = 2*\lw] (1.2,-1.4+\stag*\x)--(1.2,.4+\stag*\x);
				\draw[red,line width = 2*\lw] (0,.5+\stag*\x)--(0,-.5+\stag*\x);
	}
	
	\fill[white,path fading=  west] (-5,-\stag-.01) rectangle (5,-\stag/2);
	\fill[white] (-5,-\stag/2-\stag) rectangle (5,-\stag);
	
	\fill[white,path fading = east] (-5,2*\stag+0.01) rectangle (5,\stag + \stag/2);
	\fill[white] (-5,\stag/2+2*\stag) rectangle (5,2*\stag);

	\end{scope}

	\begin{scope}[shift = {(-.3\linewidth,0)}, scale =.5]
	\path [path fading = circle with fuzzy edge 15 percent, fill=confgrey, even odd rule] (0,0) circle (\zeroRad) (0,0) circle (1.3*\zeroRad);

	\draw[blue,fill= confcol,line width = \lw] (0,0) circle (\zeroRad);
				\draw[branchCol,dashed,line width = \lw] (0,0)--(\zeroRad,0);
					\fill[branchCol] (0,0) circle (.2);
	\draw [green,domain=160:300,line width = 2*\lw] plot ({2*cos(\x)}, {2*sin(\x)});
		\draw [magenta,domain=30:100,line width = 2*\lw] plot ({1.4*cos(\x)}, {1.4*sin(\x)});
			\draw [red,domain=120:200,line width = 2*\lw] plot ({1.6*cos(\x)}, {1.6*sin(\x)});
	\node at (.8,2) {$D_\xi$};

	\end{scope}

		\myPathTextAbove{$\xi = A_{\rm II}(\zeta; \gamma)$}{(-6.75,1.7)}{(-5.25,1.7)}{20}
	
			\myPathTextAbove{$z = \period\log(\xi)/(2\i\pi)$}{(-1.75,1.7)}{(-.25,1.7)}{20}
			\node at (0,2.5) {};
	\end{tikzpicture}
	\caption{A type II periodic parallel slit map for $M=3$.}
	\label{Fig:confIIa}
\end{figure}

We now introduce the type II circular slit map, $A_{II}$, defined in section 7.1 by \cite{Crowdy2006} as
\begin{align}
\xi =A_{\rm II}(\zeta; a_{\infty}) & \equiv  \frac{\omega(\zeta, a_{\infty})}{\omega(\zeta,1/\overline{a_{\infty}})}.%
\end{align}
This mapping relates each boundary circle to a circular slit of finite length, except from $C_0$ which is mapped to itself.  In addition, the point $a_{\infty}$ is mapped to the origin. An example of a type II circular slit map is illustrated in figure \ref{Fig:confIIa}. Since the image of every circle has constant radius, taking the scaled logarithm \eqref{Eq:log} results in a periodically repeating arrangement of slits with the image of $C_0$ forming a boundary of the period window, as illustrated in figure \ref{Fig:confIIa}.

\subsection{Type III periodic conformal slit maps} \label{Ap:conf3}

\begin{figure}[!h]
	\def\stag{3}
	\centering
	\begin{tikzpicture}[scale = .7]

	\begin{scope}[shift = {(-.7\linewidth,0)},scale =.5]
	
	\path [path fading = circle with fuzzy edge 15 percent, fill=confgrey, even odd rule] (0,0) circle (\zeroRad) (0,0) circle (1.3*\zeroRad);
	\draw[blue,fill= confcol,line width = \lw] (0,0) circle (\zeroRad);
	\node at (-.5,1.85) {$D_\zeta$};

	\skCircBlankBThick{(\delIx,\delIy)}{\radI}{red}
	\skCircBlankBThick{(\delIIx,\delIIy)}{\radII}{green}
	\skCircBlankBThick{(\delIIIx,\delIIIy)}{\radIII}{magenta}
	
	\end{scope}
	
	\begin{scope}[shift = {(.35\linewidth,0)}, rotate = 90, xscale=.4, yscale = .7]

	\foreach \x in {-1,0,1}
	{
		\draw[dashed,thick,branchCol] (-3,1+\stag*\x)--(3,1+\stag*\x);
	}
	
	\fill[confgrey, path fading = south] (-5,-3) rectangle (-3,6);
	\draw[blue, line width = \lw] (-3,-3)--(-3,6);
	
	\fill[confgrey, path fading = north] (3,6) rectangle (5,-3);
	\draw[red, line width = \lw] (3,-3)--(3,6);
	\node at (2,2.5) {$D_z$};
	
	\foreach \x in {-1,0,1,2}
	{
		\draw[green,line width = 2*\lw] (-1,-.8+\stag*\x)--(-1,.7+\stag*\x);
		\draw[magenta,line width = 2*\lw] (1.2,-1.4+\stag*\x)--(1.2,.4+\stag*\x);
	}
	
	\fill[white,path fading=  west] (-5,-\stag-.01) rectangle (5,-\stag/2);
	\fill[white] (-5,-\stag/2-\stag) rectangle (5,-\stag);
	
	\fill[white,path fading = east] (-5,2*\stag+0.01) rectangle (5,\stag + \stag/2);
	\fill[white] (-5,\stag/2+2*\stag) rectangle (5,2*\stag);

	\end{scope}

	\begin{scope}[shift = {(-.3\linewidth,0)}, scale =.5]
	\path [path fading = circle with fuzzy edge 15 percent, fill=confgrey, even odd rule] (0,0) circle (\zeroRad) (0,0) circle (1.3*\zeroRad);

	\draw[blue,fill= confcol,line width = \lw] (0,0) circle (\zeroRad);
				\draw[branchCol,dashed,line width = \lw] (0,0)--(\zeroRad,0);
	\draw[*-,thick,dashed,branchCol] (0,0);
	\draw[red, fill = confgrey, line width = \lw] (0,0 )circle (1.1);
	\draw [green,domain=160:300,line width = 2*\lw] plot ({2*cos(\x)}, {2*sin(\x)});
		\draw [magenta,domain=30:100,line width = 2*\lw] plot ({1.4*cos(\x)}, {1.4*sin(\x)});
	\node at (.8,2) {$D_\xi$};

	\end{scope}

		\myPathTextAbove{$\xi = A_{\rm III}(\zeta; \gamma)$}{(-6.75,1.7)}{(-5.25,1.7)}{20}
	
			\myPathTextAbove{$z = \period \log(\xi)/(2\i\pi)$}{(-1.75,1.7)}{(-.25,1.7)}{20}
			\node at (0,2.5) {};
	\end{tikzpicture}
	\caption{A typical type III periodic parallel slit map for $M=3$.}
	\label{Fig:circSlitIII}
\end{figure}
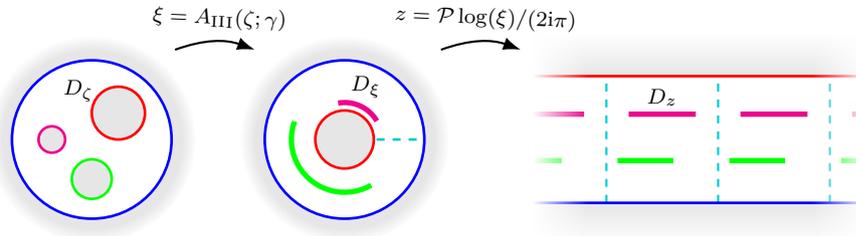

Finally, we introduce the type III circular slit map, $R_{\rm III}$, defined in section 7.3 of \cite{Crowdy2006} by
\begin{align}
\xi =A_{\rm III}(\zeta; \gamma) & = \frac{\omega(\zeta,1/\overline{\gamma})}{\omega(\zeta, \theta_1(1/\overline{\gamma})) }, \label{Eq:confIIa}
\end{align}
for any $\gamma \in D_{\zeta}$. This mapping relates each boundary circle to a circular slit of finite length except $C_0$, which is mapped to itself, and $C_1$ which is mapped to a disc centred at the origin. Taking the scaled logarithm \eqref{Eq:log} then generates a channel with periodically repeated horizontal slits, as illustrated in figure \ref{Fig:circSlitIII}.

\bibliographystyle{plain}
\bibliography{library}

\begin{thebibliography}{10}

\bibitem{Fokas2003}
M.~J. Ablowitz and A.~S. Fokas.
\newblock {\em {Complex Variables: Introduction and Applications}}, volume
  XXXIII.
\newblock Cambridge University Press, 2003.

\bibitem{Akhiezer1990}
N.~I. Akhiezer.
\newblock {Elements of the theory of elliptic functions}.
\newblock In {\em Transl. Math. Monogr.} American Mathematical Society,
  Providence, vol 79 edition, 1990.

\bibitem{Aref2007}
H.~Aref.
\newblock {Point vortex dynamics: A classical mathematics playground}.
\newblock {\em J. Math. Phys.}, 48(6):065401, 2007.

\bibitem{Ayton2018}
L.~J. Ayton.
\newblock {Analytic solution for aerodynamic noise generated by plates with
  spanwise-varying trailing edges}.
\newblock {\em J. Fluid Mech.}, 849:448--466, 2018.

\bibitem{Baddoo2018c}
P.~J. Baddoo and L.~J. Ayton.
\newblock {Potential flow through a cascade of aerofoils: direct and inverse
  problems}.
\newblock {\em Proc. R. Soc. A Math. Phys. Eng. Sci.}, 474(2217):20180065,
  2018.

\bibitem{Baddoo2020Aero}
P.~J. Baddoo and L.~J. Ayton.
\newblock {An analytic solution for gust-cascade interaction noise including
  effects of realistic aerofoil geometry [IN PRESS]}.
\newblock {\em J. Fluid Mech.}, 2020.

\bibitem{Baddoo2019c}
P.~J. Baddoo and D.~G. Crowdy.
\newblock {Periodic Schwarz-Christoffel mappings with multiple boundaries per
  period}.
\newblock {\em Proc. R. Soc. A Math. Phys. Eng. Sci.}, 475(2228), 2019.

\bibitem{Baker1897}
H.~F. Baker.
\newblock {\em {Abelian functions: Abel's theorem and the allied theory of
  theta functions}}.
\newblock Cambridge University Press, 1897.

\bibitem{Batchelor1999}
G.~K. Batchelor.
\newblock {\em {An Introduction to Fluid Dynamics}}.
\newblock Cambridge University Press, 2000.

\bibitem{Chapman1992}
A.~M. Chapman and J.~J.L. Higdon.
\newblock {Oscillatory Stokes flow in periodic porous media}.
\newblock {\em Phys. Fluids A}, 4(10):2099--2116, oct 1992.

\bibitem{Crighton1985}
David~G Crighton.
\newblock {The Kutta condition in unsteady flow}.
\newblock {\em Annu. Rev. Fluid Mech.}, 17:411--445, 1985.

\bibitem{Crowdy1999}
D.~G. Crowdy.
\newblock {Exact solutions for steady capillary waves on a fluid annulus}.
\newblock {\em J. Nonlinear Sci.}, 9(6):615--640, 1999.

\bibitem{Crowdy2000}
D.~G. Crowdy.
\newblock {A new approach to free surface Euler flows with capillarity}.
\newblock {\em Stud. Appl. Math.}, 105(1):35--58, 2000.

\bibitem{Crowdy2005b}
D.~G. Crowdy.
\newblock {The Schwarz--Christoffel mapping to bounded multiply connected
  polygonal domains}.
\newblock {\em Proc. R. Soc. A Math. Phys. Eng. Sci.}, 461(2061):2653--2678,
  2005.

\bibitem{Crowdy2006b}
D.~G. Crowdy.
\newblock {Analytical solutions for uniform potential flow past multiple
  cylinders}.
\newblock {\em Eur. J. Mech. B/Fluids}, 25(4):459--470, 2006.

\bibitem{Crowdy2007}
D.~G. Crowdy.
\newblock {Schwarz--Christoffel mappings to unbounded multiply connected
  polygonal regions}.
\newblock {\em Math. Proc. Cambridge Philos. Soc.}, 142(2):319--339, 2007.

\bibitem{Crowdy2008a}
D.~G. Crowdy.
\newblock {Explicit solution for the potential flow due to an assembly of
  stirrers in an inviscid fluid}.
\newblock {\em J. Eng. Math.}, 62(4):333--344, 2008.

\bibitem{Crowdy2008}
D.~G. Crowdy.
\newblock {The Schwarz problem in multiply connected domains and the
  Schottky–Klein prime function}.
\newblock {\em Complex Var. Elliptic Equations}, 53(3):221--236, 2008.

\bibitem{Crowdy2010}
D.~G. Crowdy.
\newblock {A new calculus for two-dimensional vortex dynamics}.
\newblock {\em Theor. Comput. Fluid Dyn.}, 24(1-4):9--24, 2010.

\bibitem{Crowdy2011}
D.~G. Crowdy.
\newblock {Frictional slip lengths for unidirectional superhydrophobic grooved
  surfaces}.
\newblock {\em Phys. Fluids}, 23(7):72001, 2011.

\bibitem{Crowdy2013a}
D.~G. Crowdy.
\newblock {Analytical formulae for source and sink flows in multiply connected
  domains}.
\newblock {\em Theor. Comput. Fluid Dyn.}, 27(1-2):1--19, 2013.

\bibitem{Crowdy2017}
D.~G. Crowdy.
\newblock {Effective slip lengths for immobilized superhydrophobic surfaces}.
\newblock {\em J. Fluid Mech.}, 825:R2, 2017.

\bibitem{CrowdyBook}
D.~G. Crowdy.
\newblock {\em {Solving problems in multiply connected domains}}.
\newblock SIAM CBMS-NSF Regional Conference Series in Applied Mathematics,
  2020.

\bibitem{Crowdy2012a}
D.~G. Crowdy, A.~S. Fokas, and C.~C. Green.
\newblock {Conformal mappings to multiply connected polycircular arc domains}.
\newblock {\em Comput. Methods Funct. Theory}, 11(2):685--706, jan 2011.

\bibitem{Crowdy2011a}
D.~G. Crowdy and C.~C. Green.
\newblock {Analytical solutions for von K{\'{a}}rm{\'{a}}n streets of hollow
  vortices}.
\newblock {\em Phys. Fluids}, 23(12):126602, 2011.

\bibitem{Crowdy2016a}
D.~G. Crowdy, E.~H. Kropf, C.~C. Green, and M.~M.~S. Nasser.
\newblock {The Schottky-Klein prime function: A theoretical and computational
  tool for applications}.
\newblock {\em IMA J. Appl. Math.}, 81(3):589--628, 2016.

\bibitem{Crowdy2005}
D.~G. Crowdy and J.~Marshall.
\newblock {Analytical formulae for the Kirchhoff-Routh path function in
  multiply connected domains}.
\newblock {\em Proc. R. Soc. A Math. Phys. Eng. Sci.}, 461(2060):2477--2501,
  2005.

\bibitem{Crowdy2006}
D.~G. Crowdy and J.~Marshall.
\newblock {Conformal mappings between canonical multiply connected domains}.
\newblock {\em Comput. Methods Funct. Theory}, 6(1):59--76, 2006.

\bibitem{Crowdy2007d}
D.~G. Crowdy and J.~S. Marshall.
\newblock {Computing the Schottky-Klein prime function on the Schottky double
  of planar domains}.
\newblock {\em Comput. Methods Funct. Theory}, 7(1):293--308, apr 2007.

\bibitem{Crowdy2010a}
D.~G. Crowdy and R.~Nelson.
\newblock {Steady interaction of a vortex street with a shear flow}.
\newblock {\em Phys. Fluids}, 22(9):096601, sep 2010.

\bibitem{Darakananda2018a}
D.~Darakananda, A.~F. de~C. da~Silva, T.~Colonius, and J.~D. Eldredge.
\newblock {Data-assimilated low-order vortex modeling of separated flows}.
\newblock {\em Phys. Rev. Fluids}, 3(12):124701, 2018.

\bibitem{Darakananda2019}
Darwin Darakananda and Jeff~D. Eldredge.
\newblock {A versatile taxonomy of low-dimensional vortex models for unsteady
  aerodynamics}.
\newblock {\em J. Fluid Mech.}, 858:917--948, 2019.

\bibitem{Delillo2004}
T.~K. DeLillo, A.~R. Elcrat, and J.~A. Pfaltzgraff.
\newblock {Schwarz-Christoffel mapping of multiply connected domains}.
\newblock {\em J. d'Analyse Math.}, 94(1):17--47, 2004.

\bibitem{Driscoll1996}
T.~A. Driscoll.
\newblock {Algorithm 756; a MATLAB toolbox for Schwarz-Christoffel mapping}.
\newblock {\em ACM Trans. Math. Softw.}, 22(2):168--186, 1996.

\bibitem{Driscoll2005}
T.~A. Driscoll.
\newblock {Algorithm 843: Improvements to the Schwarz-Christoffel toolbox for
  MATLAB}.
\newblock {\em ACM Trans. Math. Softw.}, 31(2):239--251, 2005.

\bibitem{Driscoll2002}
T.~A. Driscoll and L.~N. Trefethen.
\newblock {\em {Schwarz-Christoffel Mapping}}.
\newblock Cambridge University Press, Cambridge, 2002.

\bibitem{Ehrenstein1996}
U.~Ehrenstein.
\newblock {On the linear stability of channel flow over riblets}.
\newblock {\em Phys. Fluids}, 8(11):3194--3196, 1996.

\bibitem{Eldredge2019a}
J.~D. Eldredge.
\newblock {\em {Mathematical modeling of unsteady inviscid flows}}, volume~50
  of {\em Interdisciplinary Applied Mathematics}.
\newblock Springer International Publishing, 2019.

\bibitem{Evers2002}
I.~Evers and N.~Peake.
\newblock {On sound generation by the interaction between turbulence and a
  cascade of airfoils with non-uniform mean flow}.
\newblock {\em J. Fluid Mech.}, 463:25--52, 2002.

\bibitem{Floryan1985a}
J~M Floryan.
\newblock {Conformal-mapping-based coordinate generation method for channel
  flows}.
\newblock {\em J. Comput. Phys.}, 58(2):229--245, 1985.

\bibitem{Floryan1993}
J.~M. Floryan and C.~Zemach.
\newblock {Schwarz-Christoffel methods for conformal mapping of regions with a
  periodic boundary}.
\newblock {\em J. Comput. Appl. Math.}, 46(1-2):77--102, 1993.

\bibitem{Goodman1960}
A.~W. Goodman.
\newblock {Conformal mapping onto certain curvilinear polygons}.
\newblock {\em Univ. Nac. Tucum´an Rev. Ser. A}, 13(20):6, 1960.

\bibitem{Henrici1986}
P.~Henrici.
\newblock {\em {Applied and Computational Complex Analysis, Volume 3: Discrete
  Fourier Analysis, Cauchy Integrals, Construction of Conformal Maps, Univalent
  Functions}}.
\newblock Wiley, 1986.

\bibitem{Howe2003a}
M.~S. Howe.
\newblock {\em {Theory of vortex sound}}.
\newblock Cambridge University Press, 2003.

\bibitem{Joukowski1910}
N.~Joukowski.
\newblock {{\"{U}}ber die Konturen der Tragfl{\"{a}}chen der Drachenflieger}.
\newblock {\em Zeitschrift f{\"{u}}r Flugtechnik und Mot.}, 1:281----284, 1910.

\bibitem{Katz2009}
J.~Katz and A.~Plotkin.
\newblock {\em {Low-Speed Aerodynamics}}.
\newblock Cambridge University Press, Cambridge, 2009.

\bibitem{Kirk2018}
T.~L. Kirk.
\newblock {Asymptotic formulae for flow in superhydrophobic channels with
  longitudinal ridges and protruding menisci}.
\newblock {\em J. Fluid Mech.}, 839:R31--R312, 2018.

\bibitem{Kropf2012}
E.~Kropf.
\newblock {\em {Numerical computation of Schwarz-Christoffel transformations
  and slit maps for multiply connected domains}}.
\newblock PhD thesis, 2012.

\bibitem{Lee1976}
L.~P. Lee.
\newblock {Conformal projections based on Jacobian elliptic functions}.
\newblock {\em Cartogr. Int. J. Geogr. Inf. Geovisualization}, 13(1):67--101,
  jun 1976.

\bibitem{LlewellynSmith2011}
S.~G. {Llewellyn Smith}.
\newblock {How do singularities move in potential flow?}
\newblock {\em Phys. D Nonlinear Phenom.}, 240(20):1644--1651, 2011.

\bibitem{Peake1995}
N.~Peake and E.~J. Kerschen.
\newblock {A uniform asymptotic approximation for high-frequency unsteady
  cascade flow}.
\newblock {\em Proc. R. Soc. A Math. Phys. Eng. Sci.}, 449(1935):177--186,
  1995.

\bibitem{Peake1997}
N.~Peake and E.~J. Kerschen.
\newblock {Influence of mean loading on noise generated by the interaction of
  gusts with a flat-plate cascade: Upstream radiation}.
\newblock {\em J. Fluid Mech.}, 347:315--346, 1997.

\bibitem{Ramesh2014}
K.~Ramesh, A.~Gopalarathnam, K.~Granlund, M.~V. Ol, and J.~R. Edwards.
\newblock {Discrete-vortex method with novel shedding criterion for unsteady
  aerofoil flows with intermittent leading-edge vortex shedding}.
\newblock {\em J. Fluid Mech.}, 751:500--538, 2014.

\bibitem{Robinson1956}
A.~Robinson and J.~A. Laurmann.
\newblock {\em {Wing Theory}}.
\newblock Cambridge University Press, 1956.

\bibitem{Schmid2017}
P.~J. Schmid, M.~F. {De Pando}, and N.~Peake.
\newblock {Stability analysis for n-periodic arrays of fluid systems}.
\newblock {\em Phys. Rev. Fluids}, 2(11):113902, 2017.

\bibitem{Southwick2015}
O~R Southwick, E~R Johnson, and N.~R. McDonald.
\newblock {A point vortex model for the formation of ocean eddies by flow
  separation}.
\newblock {\em Phys. Fluids}, 27(1):16604, 2015.

\bibitem{Taira2007}
K.~Taira and T.~Colonius.
\newblock {The immersed boundary method: A projection approach}.
\newblock {\em J. Comput. Phys.}, 225(2):2118--2137, 2007.

\bibitem{Vasconcelos1993}
G.~L. Vasconcelos.
\newblock {Exact solutions for a stream of bubbles in a Hele-Shaw cell}.
\newblock {\em Proc. R. Soc. A Math. Phys. Eng. Sci.}, 442(1915):463--468,
  1993.

\bibitem{Vasconcelos2015}
G.~L Vasconcelos.
\newblock {Multiple bubbles and fingers in a Hele-Shaw channel: complete set of
  steady solutions}.
\newblock {\em J. Fluid Mech.}, 780:299--326, 2015.

\end{thebibliography}

\end{document}